\documentclass{aa}  
\usepackage{graphicx}
\usepackage{hyperref}
\usepackage{txfonts}
\begin{document}
   \title{Parameters of Herbig Ae/Be and Vega-type stars\thanks{Based
       on observations collected at the German-Spanish Astronomical Centre,
       Calar Alto, jointly operated by the Max-Planck Institut f\"ur
       Astronomie and the Instituto de Astrof\'{\i}sica de Andaluc\'{\i}a
       (CSIC), and on observations made with the WHT telescope operated on the
       island of La Palma by the Isaac Newton Group in the Spanish
       Observatorio del Roque de los Muchachos of the Instituto de
       Astrof\'{\i}sica de Canarias.}}

   \author{B. Montesinos\inst{1,2}
          \and
          C. Eiroa\inst{3}
          \and
          A. Mora\inst{3}
          \and
          B. Mer\'{\i}n\inst{4}
          }

   \offprints{Benjam\'{\i}n Montesinos\\
              \email{Benjamin.Montesinos@laeff.inta.es}}

   \institute{Instituto de Astrof\'{\i}sica de Andaluc\'{\i}a (IAA--CSIC), 
              Apartado 3004, E-18080 Granada, Spain
         \and
              Laboratorio de Astrof\'{\i}sica Espacial y F\'{\i}sica 
              Fundamental (LAEFF--INTA), European Space Astronomy Centre, 
              Apartado 78, E-28691 Villanueva de la Ca\~nada, Madrid, Spain
         \and
              Universidad Aut\'onoma de Madrid,  Departamento de F\'{\i}sica
              Te\'orica C-XI, 28049 Madrid, Spain
         \and
              Research and Scientific Support Department, European Space
              Agency (ESTEC), PO Box 299, 2200 AG Noordwijk, The Netherlands
             }

   \date{Received July 16, 2008; revised October 23, 2008 ; accepted November
   13, 2008}

 
  \abstract
   {This work presents the characterization of 27 young early-type stars,
  most of them in the age range 1--10 Myr, and three --suspected-- hot 
  companions of post-T Tauri stars belonging to the Lindroos binary sample. 
  Most of these objects show IR excesses in their spectral energy
  distributions, which are indicative of the presence of disks. The work is 
  relevant in the fields of stellar physics, physics of disks and formation 
  of planetary systems.}
   {The aim of the work is the determination of the effective temperature,
  gravity, metallicity, mass, luminosity and age of these stars. An accurate
  modelling of their disks needs, as a previous step, the knowledge of most of
  these parameters, since they will determine the energy input received by the 
  disk and hence, its geometry and global properties.}
   {Spectral energy distributions and mid-resolution spectra were used to
  estimate $T_{\rm eff}$, the effective temperature. The comparison of the
  profiles of the Balmer lines with synthetic profiles provides the value of
  the stellar gravity, $g_*$. High-resolution optical observations and
  synthetic spectra are used to estimate the metallicity, [M/H]. Once $T_{\rm
  eff}$, $g_*$ and [M/H] are known for each star, evolutionary tracks and
  isochrones provide estimations of the mass, luminosity, age and distances
  (or upper limits in some cases). The method is original in the sense that it
  is distance-independent, i.e. the estimation of the stellar parameters does
  not require, as it happens in other works, the knowledge of the distance to
  the object.}
   {Stellar parameters (effective temperature, gravity, metallicity, mass,
  luminosity, age and distances --or upper limits) are obtained for the
  sample of stars mentioned above. A detailed discussion on some individual
  objects, in particular VV Ser, RR Tau, 49 Cet and the three suspected hot
  companions of  post-T Tauris, is presented.}        
   {These results, apart from their intrinsic interest, would be extremely
  valuable to proceed a step further and attempt to model the disks
  surrounding the stars. The paper also shows the difficulty posed by the
  morphology and behaviour of the system star+disk in the computation of the
  stellar parameters.}

   \keywords{Stars: pre-main sequence -- Stars: fundamental parameters --
   Stars: abundances -- Stars: evolution -- Stars: circumstellar matter --
   Stars: planetary systems: protoplanetary disks}

   \maketitle
%

\section{Introduction}

One of the most active research areas in PMS (pre-main sequence) stellar
physics is the study of stars with protoplanetary disks. Today it is widely
accepted that circumstellar disks are a common feature of the star formation
processes (Beckwith, 1996).  The percentage of the presence of accretion
disks around young stars ranges from 50 to 100\%, at least for the youngest
star-forming regions (Kenyon \& Hartmann, 1995; Hillenbrand et al. 1998). A
study in the near infrared of young clusters by Haisch et al. (2001) in the
age range 0.3--30 Myr shows that at the early ages, the cluster disk fraction
is around 80\%, and decreases as the cluster age increases, the typical
overall disk lifetime being $5-7$ Myr. The work by those authors,
confirmed by Hillenbrand (2005), Cieza (2008) and Hern\'andez et al. (2008)
was based on samples containing mostly T Tauri stars (spectral types K and
M). In the mass range of the Herbig Ae/Be stars (spectral types B to F),
Hern\'andez et al. (2005) found that the lifetime of primordial disks is less
than $\sim\!3$ Myr. There is evidence that the typical times for disk
dissipation depend on stellar mass (e.g. Sicilia-Aguilar et al. 2005),
although in all cases this phenomenon seems to occur in the first few million
years of the stellar lifetime.

The formation of planets and planetary systems takes place in circumstellar
disks. The report in 1995 of the discovery of the first extrasolar planet
around a solar-type star, namely 51 Peg, by Mayor \& Queloz (1995), and the
subsequent detections of candidate planets increased even further the
interest in that area, and in particular, the study of the evolution of PMS
stars and their disks in a range of ages of the order of $\sim\!10^6$ years,
when the disks have their characteristic morphology and processes of potential
formation of planetesimals have already been triggered.

The mass, radius and effective temperature of the central star will determine
the energy received by the disk and therefore --at least in part-- its
geometry, energy balance, accretion rate and contribution to the spectral
energy distribution (SED hereafter). The metal abundance of the star, a
parameter which is many times overlooked and assumed to take the solar value,
is also an important factor because the gas contained in the disk will have,
in a first approximation, a similar metal content, which, in turn, will have
an influence on its evolution and the potential formation of planets (see
e.g. Wyatt et al. 2007, and references therein). In addition, the gas
dominates the mass and dynamics of the disk (e.g. Hillenbrand, 2008). The
knowledge of the metallicity is also important to place the star in the
appropriate set of evolutionary tracks to determine parameters such as mass,
luminosity and age (Mer\'{\i}n et al., 2004). Therefore, in this context, the
determination of parameters and properties of stars --i.e. what we call their
`characterization'-- in the critical age range of 1--10 Myr, is a necessary
step to model accurately the disks.

Several works have aimed at a systematic determination of parameters of PMS
and young MS stars (T Tauri, Herbig Ae/Be, Vega and A-shell) with
disks. Dunkin et al. (1997a,b) carried out a complete study of 14 Vega-type
objects based on high-resolution spectroscopy.  Kovalchuk \& Pugach (1997)
determined the gravity for a sample of 19 Herbig Ae/Be stars.  The spectral
classification, projected rotational velocities and variability of 80 PMS and
young MS stars were studied by Eiroa et al. (2001), Eiroa et al. (2002), Mora
et al. (2001) and Oudmaijer et al. (2001). Hern\'andez et al. (2004) carried
out an analysis of 75 Herbig Ae/Be stars, deriving spectral types and studying
the properties of the extinction caused by the environment of these objects.
Acke \& Waelkens (2004) presented an abundance study of 24 Herbig Ae/Be and
Vega-type stars searching for the $\lambda$ Bootis phenomenon, and Guimar\~aes
et al. (2006) determined stellar gravities and metal abundances for 12 Herbig
Ae/Be stars. Manoj et al. (2006) studied the evolution of the emission line
activity for a sample of 45 Herbig Ae/Be stars and estimated masses,
luminosities and ages for an enlarged sample of 91 objects of this type.

In this paper we present the results of the determination of effective
temperatures, stellar gravities, metallicities, masses, luminosities, ages and
distances for 30 young stars. In Sections \ref{Section:Sample} and
\ref{Section:Observations} we describe the sample of stars and the
observations. Section \ref{Section:Computation} describes how the stellar
parameters have been estimated.  Sections \ref{Section:Results} and
\ref{Section:Discussion} present the results and a detailed discussion.
Finally, some remarks are drafted in Section \ref{Section:Remarks}.

\section{The sample of stars}
\label{Section:Sample}

In this work we study the properties of a sample of 30 early-type objects,
namely 19 Herbig Ae/Be (hereafter HAeBe), five Vega-type, three A-shell and
three suspected hot companions of post-T Tauri stars (see subsection
\ref{Subsection:Individual} to see why we add the qualifier `suspected' to
these stars). In Table \ref{Table:Sample} we list some properties of the
stars. Column 1 gives the identification of the object, column 2 provides
other identifications, when available, from the MWC, BD or HD catalogues;
column 3, the type of star; column 4, the projected rotational velocity $v
\sin i$ and column 5 indicates whether the star has emission in the H$\alpha$
line or not. Since the presence or absence of this emission is an indicator of
activity, these data will be useful to discuss the evolutionary status of
the objects. The stars have been chosen from the EXPORT sample (Eiroa et al. 2000).

\begin{table}[th]
  \caption[]{The sample of stars.}
  \begin{tabular}{llrlc}
    \hline\hline
Star & Other ID & Type of star & $v \sin i$ (km/s) & H$\alpha$ \\        
    \hline
HD 31648        & MWC 480      & HAeBe     & 102 $\pm$ 5   &   Y\\
HD 58647        & BD$-13$ 2008 & HAeBe     & 118 $\pm$ 4   &   Y\\
HD 150193       & MWC 863      & HAeBe     &               &   Y\\
HD 163296       & MWC 275      & HAeBe     & 133 $\pm$ 6   &   Y\\
HD 179218       & MWC 614      & HAeBe     &               &   Y\\
HD 190073       & MWC 325      & HAeBe     &               &   Y\\
HR 10           & HD 256       & Ash       & 294 $\pm$ 9   &   N\\
HR 26 A         & HD 560 A     & HPTT      &               &   N\\
HR 2174         & HD 4211      & Vega      & 252 $\pm$ 7   &   N\\
HR 4757 A       & HD 108767 A  & HPTT      & 239 $\pm$ 7   &   N\\
HR 5422 A       & HD 127304 A  & HPTT      & 7.4 $\pm$ 0.3 &   N\\
HR 9043         & HD 223884    & Vega      & 205 $\pm$ 18  &   N\\
AS 442          &              & HAeBe     &               &   Y\\
BD+31 643       & HD 281159    & Vega      & 162 $\pm$ 13  &   N\\
$\lambda$ Boo   & HD 125162    & Vega      & 129 $\pm$ 7   &   N\\
VX Cas          &              & HAeBe     & 179 $\pm$ 18  &   Y\\
SV Cep          & BD+72 1031   & HAeBe     & 206 $\pm$ 13  &   Y\\
49 Cet          & HD 9672      & Vega      & 186 $\pm$ 4   &   N\\
24 CVn          & HD 118232    & Ash       & 173 $\pm$ 4   &   N\\
51 Oph          & HD 158643    & HAeBe     & 256 $\pm$ 11  &   Y\\
T Ori           & BD$-05$ 1329 & HAeBe     & 175 $\pm$ 14  &   Y\\
BF Ori          & BD$-06$ 1259 & HAeBe     & 37 $\pm$ 2    &   Y\\
UX Ori          & HD 293782    & HAeBe     & 215 $\pm$ 15  &   Y\\
V346 Ori        & HD 287841    & HAeBe     &               &   Y\\
V350 Ori        &              & HAeBe     &               &   Y\\
XY Per          & HD 275877    & HAeBe     & 217 $\pm$ 13  &   Y\\
VV Ser          &              & HAeBe     & 229 $\pm$ 9   &   Y\\
17 Sex          & HD 88195     & Ash       & 259 $\pm$ 13  &   N\\
RR Tau          & BD+26 887a   & HAeBe     & 225 $\pm$ 35  &   Y\\
WW Vul          & HD 344361    & HAeBe     & 220 $\pm$ 22  &   Y\\   
    \hline
\end{tabular}

\begin{minipage}{8.5cm}
  \underline{Notes to Table \ref{Table:Sample}}: The abbreviations in
  column 3 mean: HAeBe (Herbig Ae/Be), HPTT (hot companion of a post-T Tauri
  star) and Ash (A-shell star).

  The data for $v \sin i$ are from Mora et al. (2001). Slight refinements 
  to the results for SV Cep (225 km/s), XY Per (200 km/s) and WW Vul (210
  km/s) were presented by Mora et al. (2004). A blank in column 4
  means that the star was not observed in high resolution, furthermore the
  determination of this parameter was not feasible and its value has not
  been found elsewhere.
\end{minipage}
\label{Table:Sample}
\end{table}

Apart from the intrinsic interest of the objects, one of the peculiarities of
analysing this kind of early-type objects is that the determination of
gravities, as we will see in subsection \ref{Subsection:Gravities}, is
feasible due to the fact that, for stars with $T_{\rm eff}$ above
approximately 7500 K, the shapes of the profiles of the Balmer lines have a
strong dependence with the stellar gravity.  Provided that an estimation of
$T_{\rm eff}$ is available, $\log g_*$ can be derived and then proceed with
the determination of other stellar parameters.

The stars HR 26 A, HR 4757 A and HR 5422 A, are the hot components of three
binary systems whose cool star is a post-T Tauri object. The systems belong to
the so-called Lindroos binary sample (Lindroos, 1985). We will discuss their 
evolutionary status and their physical link with the corresponding cool stars  
in Section \ref{Subsection:Individual}.

\section{The observations}
\label{Section:Observations}

Two sets of observations have been used in this work. Intermediate resolution,
high signal-to-noise spectra in the range 3700--6200 \AA, were obtained for
all the stars with CAFOS (Calar Alto Faint Object Spectrograph) on the 2.2-m
telescope at Calar Alto Observatory (CAHA, Almer\'{\i}a, Spain). The
observations were taken in two campaigns, 28 October--2 November 2004 and 1--4
April 2005. The motivation was to obtain high quality Balmer line profiles, in
particular H$\beta$, H$\gamma$ and H$\delta$, in order to estimate the stellar
gravities (see subsection \ref{Subsection:Gravities}).  CAFOS was equipped
with a CCD SITe detector of 2048$\times$2048 pixels (pixel size 24 $\mu$m) and
the grism Blue-100, centered at 4238 \AA, giving a linear dispersion of $\sim
88$ \AA/mm (2 \AA/pixel), which is equivalent to a resolving power of
$\sim\!2500$. A slit width of 130 $\mu$m was used, corresponding to 1.5 arcsec
projected on the sky.  The usual bias, dark and dome flat-field frames were
taken each night. Standard procedures were used to process the data. The
nominal positions of the Balmer lines were used to self-calibrate the spectra
in wavelength; a second-order polynomial wavelength-pixel was used. The widths
of the lines of the calibration Hg-He-Rb lamps were used to obtain the
width of the instrumental profile of the spectrograph for each night.

\begin{figure}
\centering
\includegraphics[width=9cm]{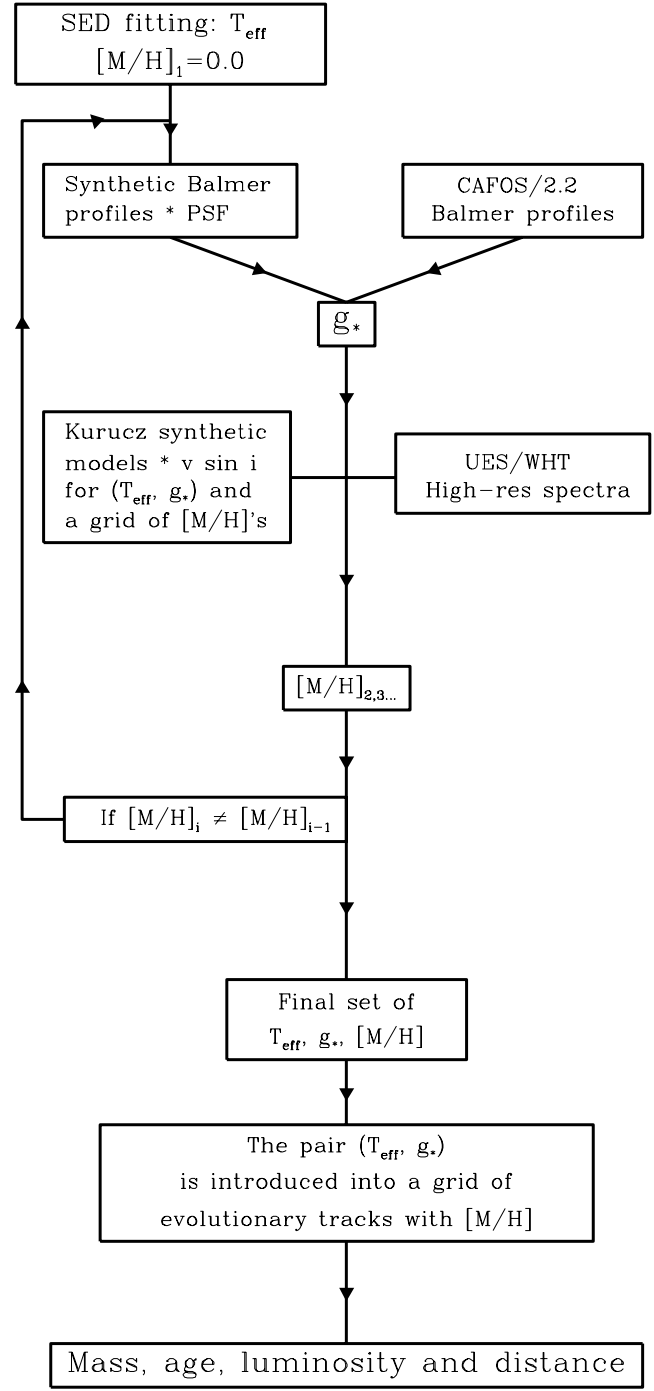}
 \caption{A flow chart sketching the procedure we use to estimate 
          the stellar parameters.}
 \label{Figure:flowchart}
\end{figure}

High-resolution echelle spectra were used in the computation of the
metallicities (subsection \ref{Subsection:Metals}). These spectra were taken
during the four EXPORT campaigns in 1998 and 1999 with the Utrech Echelle
Spectrograph (UES) on the 4.2-m William Herschel Telescope (WHT) at the
Observatorio del Roque de los Muchachos (La Palma, Spain). UES was set up to
provide a wavelength coverage between 3800 and 5900 \AA. The spectra were
dispersed into 59 echelle orders with a resolving power of 48\,000.  The slit
width was set to 1.15 arcsec projected on the sky. Details on the observing
runs and the reduction of the UES observations can be found in Mora et al.
(2001).

\section{The computation of the stellar parameters}
\label{Section:Computation}

Fig. \ref{Figure:flowchart} shows a sketch of the whole procedure we follow to
compute the stellar parameters. In the next paragraphs of this Section we
describe in detail every step.

\subsection{The effective temperature}
\label{Subsection:Teffective}

The effective temperature is the first quantity needed to start the
computation of the whole set of parameters. The procedure to estimate $T_{\rm
eff}$ followed in this work and also by Mer\'{\i}n (2004), consists in a
comparison of the observed spectral energy distribution, more precisely of its
photospheric part (see subsection \ref{Subsection:Temperatures} for a
discussion on this), with a grid of low-resolution synthetic spectra (Kurucz,
1993), with different effective temperatures\footnote{At this step, the value
of the gravity of the synthetic models does not have any noticeable effect in
the estimation of the effective temperature.} and reddened with a range of
extinctions, typically from $E(B\!-\!V)=0.1$ to 1.2, in those cases when no
solution with the original models was found. The total-to-selective extinction
$R_V$ used is 3.1 although some studies (e.g. Hern\'andez et al. 2004) suggest
that a higher value ($R_V\!\sim\!5$) could be more appropriate to model the
extinction caused by the environment of HAeBe stars. We will discuss this
choice in subsection \ref{Subsection:Individual}. The synthetic spectra were
normalized to the observed SED and the best fits were chosen after a careful
comparison and assessment of each case.

In some cases a degeneracy appeared, i.e. two or more synthetic spectra fit
reasonably well the observed SED, therefore additional information (e.g. the
best fit between the observed and synthetic Balmer profiles) had to be used to
break it and choose the correct solution. In subsection
\ref{Subsection:Individual} we show how to solve this and other problems found
at this very first step for some of the objects, in particular the very
complex cases of VV Ser and RR Tau. The degeneracy cannot be broken using the
intrinsic photometric colours of standard stars compared with those of the
target stars: two synthetic spectra corresponding to two different sets of
values ($T_{\rm eff}$, $g_*$) can fit the observed photometric points for two
different values of $E(B\!-\!V)$, both of them being consistent with the
intrinsic colours of stars of each spectral type reddened to match the
observed colours.

\begin{figure*}[t]
   \centering
   \includegraphics[width=18cm]{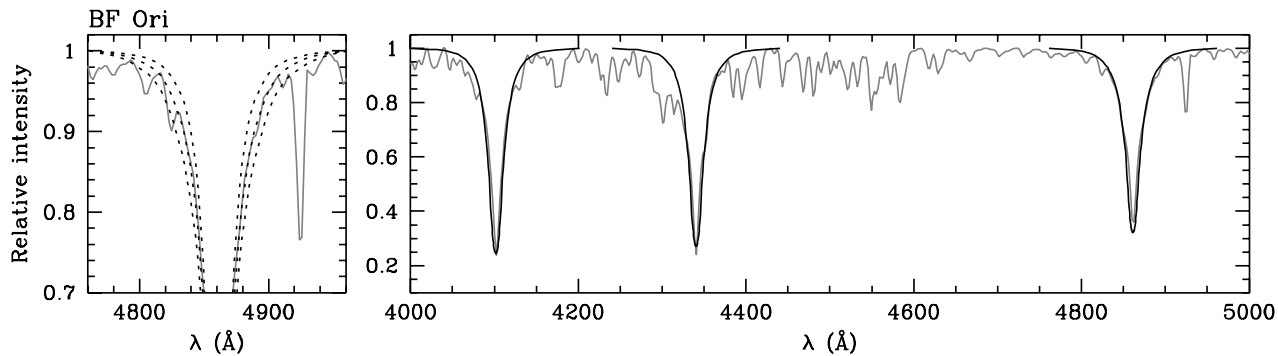}
   \caption{{\it Left}: H$\beta$ profile of BF Ori (grey solid line) and synthetic
     profiles (black dotted lines) computed with $T_{\rm eff}=8970$ K, the
     effective temperature of this star, and $\log g_*=$3.5, 4.0 and 4.5 (the
     latter profile is the broadest).  {\it Right}: Spectrum of BF Ori obtained with
     CAFOS on the 2.2-m telescope at CAHA (grey solid line) and the synthetic
     profiles for H$\delta$, H$\gamma$ and H$\beta$ computed for $T_{\rm
     eff}=8970$ K and $\log g_*=3.83$ (black solid lines) broadened with the
     instrumental profile. All the synthetic models shown in this figure were
     computed with [M/H]=+0.20, which is the metallicity estimated for this
     star.}
    \label{Figure:bfori_g}
\end{figure*}

The SEDs are built from simultaneous optical and near-infrared (near-IR)
photometry obtained in 1998 and 1999 (Eiroa et al. 2001; Oudmaijer et
al. 2001). Simultaneity of the observations is a key point at this step, since
for highly variable objects, the use of photometry from different epochs could
lead to wrong results. Among our data, the photometry corresponding to the
maximum of the stellar brightness, i.e. obtained when the star is likely to be
less obscured, is used in building the SED. We work here under the
hypothesis that the obscuration is caused by the disk, and hence, that the
brightest set of magnitudes in the optical and near-IR --up to the wavelength
where the thermal emission from the disk starts-- represents the stellar
photosphere.

Some stars were observed in the range 1200--3200 \AA{} by the International
Ultraviolet Explorer ({\it IUE}) space observatory; these
observations\footnote{{\tt http://sdc.laeff.inta.es/ines/}} are used with
caution to complete the SEDs, due to the variable nature of most of the
objects, the non-simultaneity with the optical and near-IR
observations\footnote{{\it IUE} was operative between 1978 and 1996, see Kondo
(1989) and Wamsteker \& Gonz\'alez-Riestra (1998) for details.}. In
addition to the method outlined in the first paragraph of this subsection,
the comparison between the unreddened synthetic spectra and the ultraviolet
part of the SED is particularly useful to have a better estimation of the
extinction. AAVSO\footnote{American Association of Variable Star Observers,\\
{\tt http://www.aavso.org/}} magnitudes, when they had been measured almost
simultaneously with the {\it IUE} observations, are used to place at the same
level the ultraviolet and the optical/near-IR parts of the SED. 

Table \ref{Table:Results} gives the results for the effective temperatures
found in this work and adopted throughout. In a few cases, specified at the
table, values of the effective temperature from other sources have been
used. A typical upper limit of the uncertainty in the determination of this
parameter is of the order of $\pm 200$ K.

\subsection{Stellar gravities}
\label{Subsection:Gravities}

Stellar gravities are estimated by comparing the wings of the Balmer lines
H$\beta$, H$\gamma$ and H$\delta$ with synthetic profiles extracted from
Kurucz (1993) model atmospheres. The gravity is obtained by measuring the
wavelengths of the intensity levels $I\!=\!0.80$ and $I\!=\!0.90$ --the latter
only if the profile is clean enough-- below the normalized continuum on the
blue and red sides of the observed Balmer profiles. These intensities were
chosen to avoid potential uncertainties in the overall shape of the wings near the
continuum arising both from the normalization of the original spectra and from
the presence of weak absorption lines in this region.  

A pair of values, namely $\lambda$(blue) and $\lambda$(red), is obtained for
each line for the intensity levels mentioned above. Each wavelength is then
compared with the corresponding one measured on a grid of synthetic profiles
computed for the effective temperature of the star, and different
gravities. The initial metallicity adopted is [M/H]=0.0, i.e. solar. The
synthetic profiles are convolved with the instrumental response function of
the corresponding night. The effect of rotational broadening, even for values
of $v \sin i$ as high as $\sim\!250$ km/s, is negligible in the upper part
(above $I\!\simeq\!0.50$) of the synthetic Balmer profiles, therefore this
broadening was not applied. Linear interpolation was used to convert the
discrete grid of wavelengths from the Kurucz models into a continuous
functions of $\log g_*$.  The final result for $\log g_*$ is the mean value of
the results for each line with an uncertainty corresponding to the standard
deviation $\sigma$. The observed Balmer profiles showing emission or strange
features due to the presence of several components of different widths were
not used for the computations of gravities.

In Fig. \ref{Figure:bfori_g} we see an example of the determination of the
stellar gravity. The panel on the left shows the H$\beta$ line profile (solid
line plotted in grey) for BF Ori ($T_{\rm eff}$=8970 K) and three synthetic
profiles (dotted black lines) corresponding to that effective temperature and
values of $\log g_*$ of 3.5, 4.0 and 4.5 (the larger the gravity, the broader
the synthetic profile). The linear interpolation done at the blue and red
wings of each profile and the subsequent average for the three Balmer lines
give a value of $\log g_*\!=\!3.83\pm 0.12$. The panel on the right shows the
CAFOS spectrum for this star in the interval 4000-5000 \AA, and the three
synthetic Balmer profiles for H$\delta$, H$\gamma$ and H$\beta$ computed for
$T_{\rm eff}$=8970 K and $\log g_*=3.83$. In this particular case, the metal
abundance we found was slightly larger than solar, namely [M/H]=$+0.20\pm0.05$
(see below in subsection \ref{Subsection:Metals} and Table
\ref{Table:Results}). An iteration in the estimation of the gravity following
the same process, but using that value of the metallicity in the synthesis of
the Balmer lines, gives for this star the same result (see
Fig. \ref{Figure:flowchart}). All the synthetic profiles shown in
Fig. \ref{Figure:bfori_g} were computed with [M/H]=+0.20.

\subsection{Metallicities}
\label{Subsection:Metals}

\begin{figure}[t]
   \centering
   \includegraphics[width=7.9cm]{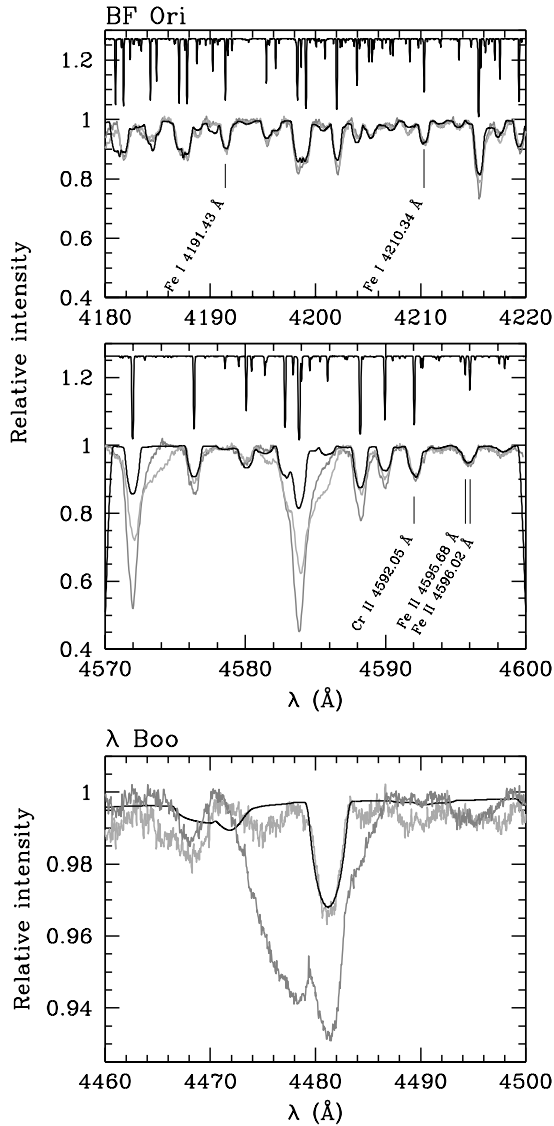}
   \caption{{\it Top}: Two sections of the UES/WHT spectra of BF Ori along
     with the synthetic spectrum. The spectra plotted in dark-grey and
     light-grey solid lines --both of them with their continuum normalized to
     an intensity of 1.0-- were obtained on 25 October 1998 and 30 January
     1999, respectively. The synthetic spectrum, computed for $T_{\rm
     eff}=8970$ K, $\log g_*=3.83$ and [M/H]=+0.20 and broadened with a
     rotation profile $v \sin i=38$ km/s has been plotted as a black solid
     line overimposed to the observed spectra. In the upper part of each panel
     we show the synthetic unbroadened spectrum --not to scale-- for
     comparison. {\it Bottom}: Two spectra of $\lambda$ Boo around the Mg~{\sc
     ii} doublet obtained on 16 May 1998 (light grey) and 29 July 1998 (dark
     grey). The first one is best fitted with [Mg/H]=$-2.0$, whereas the
     latter shows traces of high circumstellar activity. See text for further
     details.}
    \label{Figure:bforilambdaori}
\end{figure}

Metal abundances are determined by comparing the profiles of lines (or blends)
extracted from the high-resolution spectra, with synthetic spectra computed
using the ATLAS9 and SYNTHE codes by Kurucz (1993).  In this work
we have used the GNU Linux version of the codes prepared by Luca Sbordone,
Piercarlo Bonifacio and Fiorella Castelli, available
online\footnote{{\tt http://wwwuser.oat.ts.astro.it/atmos/}} (Sbordone et
al. 2004).

Model atmospheres provided by Castelli \& Kurucz (2003) are used as one
of the inputs for the synthesis. These models have more accurate solar
abundances, were computed with new opacity distribution functions (ODFs) and
contain further improvements upon previous ODFs Kurucz's models (see the paper
by Castelli \& Kurucz (2003) for details).  The ATLAS9 code allows to
compute, for a given metallicity, a model atmosphere for any value of the
effective temperature and gravity from a `close' model already present in
the Castelli \& Kurucz's grids.  In fact, this step had to be done for all the
stars in our sample because the corresponding model atmospheres did not exist
for their values of $T_{\rm eff}$ and $\log g_*$.

The spectral synthesis was carried out using SYNTHE, which is actually not a
single code, but a suite of programmes. Apart from the model atmosphere --which
contains, among other things, the atomic fractions of the different elements--
the list of the atomic and molecular transitions and the spectral range to be
synthetized are also read. SYNTHE computes the excitation and ionization
populations of neutrals and ions and then produces the final spectrum. A
special module of SYNTHE computes intensities at several inclinations in the
atmosphere and uses these, along with the value of $v \sin i$ (see Table
\ref{Table:Sample}), to simulate the spectrum of a rotating star.

Spectra corresponding to the same $T_{\rm eff}$ and $\log g_*$ are computed
for different metallicities and a comparison is done with the observed
spectrum. Most of the stars in the sample have large values of the projected
rotational velocity, therefore it is impossible to analyse individual lines
and only an average value of the metallicity can be extracted. Two regions
have been explored to estimate metallicities, namely, that between 4150 and
4280 \AA{} and the one between 4450 and 4520 \AA. In a few stars, those with
low or moderate values of $v \sin i$, individual abundances for each chemical
element could be derived from the spectra, this is the case for HR 5422 A and BF
Ori, and to a less extent, that of stars with $v \sin i$ less than around 150
km/s. In this work only average abundances, even for the low-$v \sin i$
objects, were computed.

Some stars in the sample show various and complex degrees of variability;
therefore it is important, whenever it is possible, to compare all the spectra
available for a given object and select those spectral features that present
little or no variation from one epoch to another. In Figure
\ref{Figure:bforilambdaori} we show two examples of how important this can
be. The top part of the plot shows two regions of high-resolution spectra of
BF Ori. The spectra taken on 25 October 1998 (dark-grey line) and 30 January
1999 (light-grey line) are shown together with the synthetic spectrum, which
has been overplotted as a black solid line. In the upper part of each panel,
the unbroadened spectrum shows how many transitions contribute to each
blend. It is apparent in both sections plotted, especially in the second
panel, covering the interval 4570--4600 \AA, how some lines change
dramatically from one epoch to another. Other features, which have been
labelled with the identification of the most intense line in the blend, remain
almost unchanged. The panel at the bottom shows two spectra of $\lambda$ Boo
around the Mg~{\sc ii} doublet at 4481.13--4481.33 \AA{} obtained on 16 May
1998 (light grey) and 29 July 1998 (dark grey).  The first one is best fitted
with [Mg/H]=$-2.0$, whereas the latter shows traces of high circumstellar
activity. No attempt was made to obtain any value of the abundance from this
spectrum.

\begin{figure*}
   \centering
   \includegraphics[width=16cm]{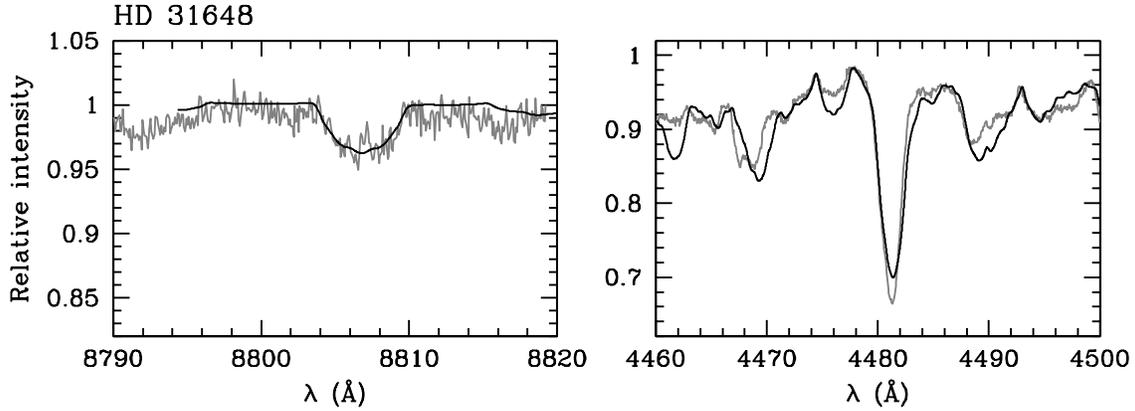}
   \caption{Two regions of the spectrum of HD 31648
     around the Mg {\sc i} line at 8806.76 \AA, and the Mg~{\sc ii}
     doublet at 4481.2 \AA, respectively (grey lines).  The Mg {\sc i}
     profile is best fitted by a model with [Mg/H]=$-0.10$, whereas the
     Mg {\sc ii} doublet is best reproduced with a synthetic model with
     [Mg/H]=$+0.50$ (black lines). }
    \label{Figure:hd31648}
\end{figure*}

Figure \ref{Figure:hd31648} is also very illustrative of another difficulty to
estimate the metal abundances for these stars: in some cases, for a given
object, lines of the same element in different ionization states, are fitted
by synthetic spectra with {\it very} different metal abundances. The plot
shows two regions of the spectrum of HD 31648, namely around the Mg {\sc i}
line at 8806.76 \AA, and around the Mg~{\sc ii} doublet at 4481.2 \AA,
respectively. The observations are not simultaneous\footnote{The spectrum of
HD 31648 around the Mg {\sc i} line was obtained in December 1996 and has been
kindly provided by Dr Bram Acke.}, the Mg {\sc i} profile is best fitted by a
model with [Mg/H]=$-0.10$, whereas the Mg {\sc ii} doublet is best reproduced
with a synthetic model with [Mg/H]=$+0.50$. Such a discrepancy could be
attributed to circumstellar contamination of the line in the latter case.

\subsection{Iterations}
\label{Subsection:Iterations}

As we mentioned at the beginning of this Section, the determination of the
effective temperature was done by comparing the SED of each star with
synthetic models with [M/H]=0.0. The first determination of the stellar
gravity, following the flow chart shown in Fig. \ref{Figure:flowchart}, also
assumed solar metallicity. The pair ($T_{\rm eff}$, $\log g_*$) was used to
extract the metallicity. When the value of the metal abundance obtained was
different by more than 0.30 dex from the solar one, we iterated the whole
process, i.e. a new effective temperature was estimated and a new gravity was
found, using in both cases models and synthetic profiles computed with the new
metallicity. It turned out that for metallicity differences as large as 0.80,
the new effective temperature did not differ from the original determination,
therefore only the gravity was recomputed for those stars. A full iteration,
i.e. the estimation of new values of $T_{\rm eff}$ and $\log g_*$, was done for
$\lambda$ Boo.  One iteration cycle was sufficient to reach the final results.

\subsection{Masses, luminosities, ages and distances}
\label{Subsection:MLAD}

As it can be seen at the bottom of Fig. \ref{Figure:flowchart}, the knowledge
of the effective temperature, gravity and metal abundance allows us to place
the star in an HR diagram $\log g_* - \log T_{\rm eff}$ and superimpose the
appropriate set of evolutionary tracks and isochrones for that specific
metallicity. This gives directly --or after a simple interpolation between the
tracks and isochrones enclosing the position of the star-- the stellar mass and
the age. In addition, since there is a one-to-one correspondence between a
pair $(T_{\rm eff},\,g_*)$ on a given track and a pair $(T_{\rm
eff},\,L_*/L_\odot)$, one can also estimate the stellar luminosity in solar
units, and if the observed --dereddened-- stellar flux, $F_*$, is known
through the SED, then, an estimation of the distance can also be done. The
values of $F_*$ are taken from Mer\'{\i}n (2004) or determinations carried out
in this work. $F_*$ is computed by fully integrating the flux of the
dereddened synthetic Kurucz model that is considered to best fit the
photospheric part of the SED.

The evolutionary tracks for a scaled solar mixture from the Yonsei-Yale group
(Yi et al.  2001) -- Y$^2$ in their notation -- have been used in this
work. The Y$^2$ set contains tracks up to $M$=5.2 M$_\odot$, therefore, the
tracks from Siess et al. (2000) for masses above that value were used in
a few cases (see subsection \ref{Subsection:ResultsMLAD} for details).

\begin{table}[th]
  \caption[]{Effective temperatures, gravities and metallicities.}

  \begin{tabular}{lrc}
\hline\hline
Star & $T_{\rm eff}$ (K)& $\log g_*$ \\
\hline
HD 58647        & 10500$^1$ &$3.33\pm0.05$ \\
HD 150193       &  8970$^1$ &$3.99\pm0.12$ \\
HD 179218       &  9500$^1$ &$3.91\pm0.07$ \\
HD 190073       &  9500$^1$ &$3.37\pm0.08$ \\
HR 26 A         & 11400$^2$ &$4.08\pm0.05$ \\ 
AS 442          & 11000$^7$ &$3.79\pm0.10$ \\
BD+31 643       & 15400$^1$ &$3.75\pm0.05$ \\
V346 Ori        &  9750$^7$ &$3.96\pm0.11$ \\
V350 Ori        &  8970$^7$ &$4.07\pm0.12$ \\
  \end{tabular}

  \begin{tabular}{lrcr}
    \hline
Star & $T_{\rm eff}$ (K) & $\log g_*$ & \multicolumn{1}{c}{[M/H]}     \\        
    \hline
HD 31648        &  8250$^7$      &$4.00\pm0.20$ &$0.00\pm0.05$     \\
HD 163296       &  9250$^7$      &$4.07\pm0.09$ &$+0.20\pm0.10$    \\
HR 10           &  9750$^7$      &$3.72\pm0.10$ &$+0.20$:          \\
HR 2174         &  8970$^1$      &$3.08\pm0.12$ &$0.00\pm0.10$     \\
HR 4757 A       &  10400$^{2,7}$ &$4.06\pm0.05$ &$+0.10\pm0.10$    \\
HR 5422 A       &  10250$^7$     &$4.08\pm0.06$ &$+0.20\pm0.10$    \\
HR 9043         &  7570$^3$      &$3.85\pm0.10$ &$0.00\pm0.10$     \\
$\lambda$ Boo   &  8750$^4$      &$4.10\pm0.12$ &$-1.50\pm0.20$    \\
VX Cas          &  10000$^7$     &$4.27\pm0.05$ &$0.00\pm0.10$     \\
SV Cep          & 10250$^{5,7}$  &$4.25\pm0.10$ &$+0.10\pm0.10$    \\
49 Cet          &  9500$^7$      &$4.30\pm0.11$ &$+0.10\pm0.10$    \\
24 CVn          &  9000$^7$      &$3.81\pm0.13$ &$+0.80\pm0.10$    \\
51 Oph          &  10250$^7$     &$3.57\pm0.06$ &$+0.10\pm0.10$    \\
T Ori           &  9750$^7$      &$4.03\pm0.10$ &$+0.10\pm0.10$    \\
BF Ori          &  8970$^1$      &$3.83\pm0.12$ &$+0.20\pm0.05$    \\
UX Ori          &  8460$^1$      &$3.89\pm0.15$ &$0.00\pm0.10$     \\
XY Per          &  9750$^7$      &$3.86\pm0.16$ &$+0.80$:          \\
VV Ser          & 13800$^{6,7}$  &$4.02\pm0.15$ &$0.00$:           \\
17 Sex          &  9520$^1$      &$3.19\pm0.10$ &$+0.10\pm0.10$    \\
RR Tau          &  10000$^7$     &$3.31\pm0.13$ &$+0.10\pm0.05$    \\
WW Vul          &  8970$^1$      &$3.90\pm0.20$ &$+0.50\pm0.10$    \\
\hline
\end{tabular}

\begin{minipage}{8.5cm}
  \underline{Notes to Table \ref{Table:Results}}: The effective temperatures
  are from $^1$Mer\'{\i}n (2004), $^2$Gerbaldi et al. (2001), $^3$Paunzen et
  al. (2006), $^4$adopted as a mean of the values proposed by Venn \& Lambert
  (1990) and Adelman et al. (2002), $^5$Mora et al. (2004), $^6$Hern\'andez et
  al. (2004), $^7$this work.
\end{minipage}

\label{Table:Results}
\end{table}      

\section{Results}
\label{Section:Results}

\subsection{Gravities and metallicities}

\begin{figure*}
   \centering
   \includegraphics[width=17cm]{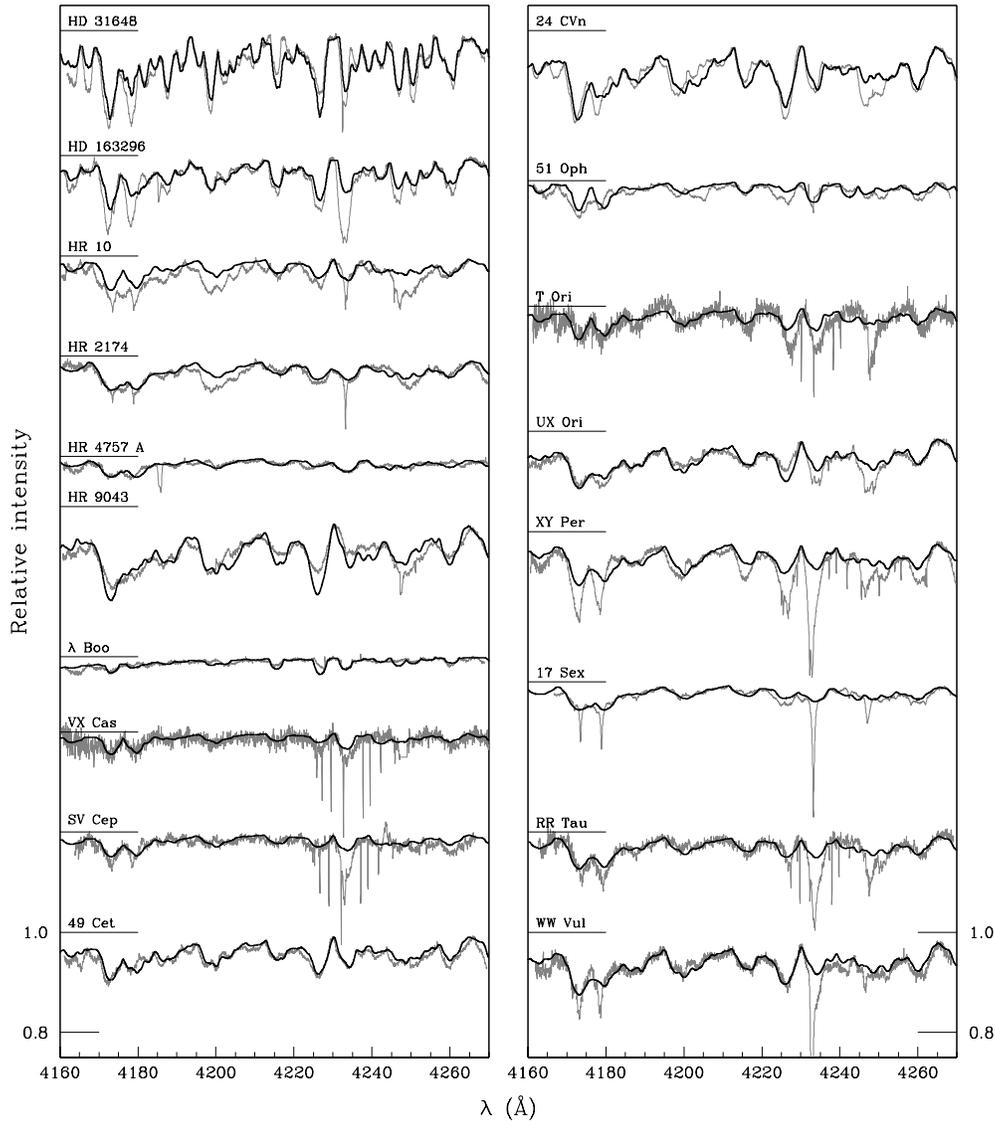}
   \caption{Selected observed spectra of the stars studied in this paper (grey lines)
     and synthetic spectra (black lines overplotted on each spectrum) computed
     with the metallicities [M/H] given in Table \ref{Table:Results}. The
     spectral range is 4160-4270 \AA.  All the spectra have been normalized
     and plotted on the same scale, and have been shifted upwards with respect
     to the lowest ones.  The small horizontal bars below the identification
     of the stars mark the intensity level unity for each pair of spectra. To
     have an idea of the range of intensities each spectrum covers, we have
     indicated the levels $I\!=\!0.8$ and $I\!=\!1.0$ corresponding to 49 Cet
     and WW Vul in both panels. The spectra of HR 5422 A, BF Ori and VV Ser are not
     shown; the first two stars have low values of $v\sin i$, therefore the
     lines are extremely narrow and the plots would look clumsy, the latter has
     a rather noisy spectrum and the resuls are uncertain. Two sections of the spectrum 
     of BF Ori and the corresponding synthetic spectrum are shown in 
     Fig. \ref{Figure:bforilambdaori}.}
\label{Figure:panel}
\end{figure*}

Table \ref{Table:Results} shows the values of the effective temperature,
gravity and metallicity (the latter when high-resolution observations were
available) for the stars in the sample. In three cases, namely HR 26 A,
HR 9043 and $\lambda$ Boo, specified in the table, values of $T_{\rm eff}$
obtained by other authors with a methodology different from that of Mer\'{\i}n
(2004) or this work have been taken; for the first two stars the calibrations
used in those papers gave more accurate results than ours, whereas for
$\lambda$ Boo the value adopted is a mean of the effective temperatures from
two works where detailed studies of this star are done. The fourth column of
the lower part of the table gives a generic value of [M/H], estimated by
comparing the observed spectra with the synthetic ones and trying to mach the
shape and depth of the non-variable blends composed of many broadened
individual lines. The determinations of [M/H] for HR 10, XY Per and VV Ser are
marked with a colon, indicating that they are very uncertain given the large
amount of circumstellar lines contaminating the photospheric spectra, or the
low signal-to-noise ratio in the spectra of the latter object; therefore these
values should be taken with caution.

In Fig. \ref{Figure:panel} we show selected spectra (grey lines) of all the
stars in the sample, apart from a few exceptions, for which an estimation of
the metal abundance has been possible and the synthetic spectra (black lines)
computed using the values of $T_{\rm eff}$, $\log g_*$ and [M/H] given in
Table \ref{Table:Results}. All the spectra are plotted on the same scale. To
show all together in the same graph we have shifted them upwards by different
amounts with respect to the lowest one in each panel. The short horizontal
bars mark where the intensity unity is placed. Each observed spectrum has been
normalized to its corresponding synthetic spectrum. The fits to HR
10 and XY Per are just tentative, showing how difficult is to find a region in
the spectra not contaminated by circumstellar lines.

\subsection{Masses, luminosities, ages and distances}
\label{Subsection:ResultsMLAD}

\begin{figure*}

   \centering
   \includegraphics[width=18cm]{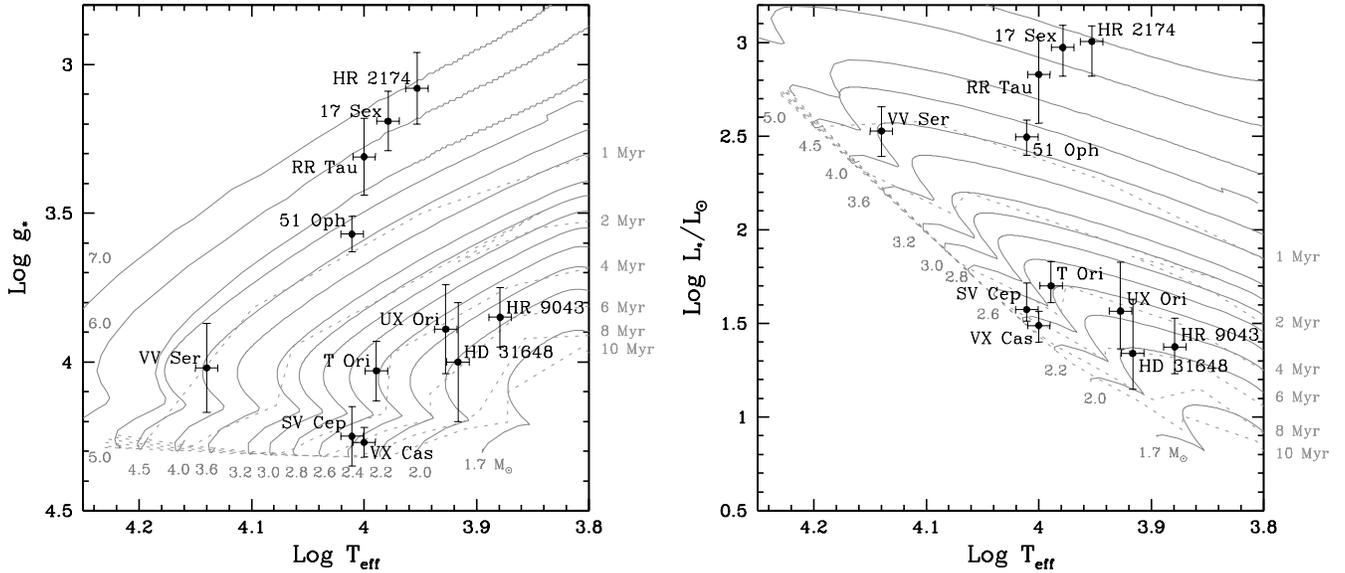}
   \caption{Left: $\log g_* - \log T_{\rm eff}$ HR diagram showing the
            position of the stars with metallicity [M/H]= 0.00 and
            $+0.10$. Right: The corresponding $\log L_*/L_\odot - \log T_{\rm
            eff}$ HR diagram for the same stars. There is a one-to-one
            correspondence between a point $(\log T_{\rm eff}, \log g_*)$ and
            a point $(\log T_{\rm eff}, \log L_*/L_\odot)$.  This allows the
            estimation of masses, ages and luminosities provided the effective
            temperature and gravity are known. The sets of PMS tracks (solid
            lines) and isochrones (dashed lines) for solar metallicity from Yi
            et al. (2001; masses up to 4.5 $M_\odot$) and Siess et al. (2000;
            5.0, 6.0 and 7.0 $M_\odot$) have been used in both plots. The
            stars HR 4757 A and 49 Cet are not included in this plot. Their
            position in the HR diagram is discussed in subsection
            \ref{Subsection:Individual}.  See text for details.}
    \label{Figure:hr}
\end{figure*}

In Table \ref{Table:MLAD} we give the results of the determination of masses,
luminosities, ages and distances, following the procedure illustrated in
Fig. \ref{Figure:flowchart} and outlined in subsection \ref{Subsection:MLAD}.
The uncertainties for each value correspond to the propagation of the errors
$\pm\sigma(\log g_*)$\footnote{Strictly speaking, one should make a
computation of the total uncertainties in the parameters considering the
combined effect of the errors {\it both} in effective temperature and
gravity. For the sake of simplicity we have only propagated the latter ones,
which are the largest on average. In Table 4 and Fig. 7 of the paper by
Mer\'{\i}n et al. (2004) we can see in detail a calculation considering both
uncertainties at a time.}. Note that in some special cases, e.g. the masses of
VV Ser or T Ori, these uncertainties can have the {\it same} sign, if the
position in the HR diagram is such that the points $(\log T_{\rm eff}, \log
g_* + \sigma(\log g_*))$ and $(\log T_{\rm eff}, \log g_* - \sigma(\log g_*))$
fall on tracks with masses both larger (or smaller) than those corresponding
to the point $(\log T_{\rm eff}, \log g_*)$ itself (see
Fig. \ref{Figure:hr}). For the nine stars at the upper part of Table
\ref{Table:Results}, for which the computation of metal abundances was not
feasible, tracks with solar metallicity have been used.  The second column of
the table lists the values of $F_*$ used to estimate the distance from the
stellar luminosity. The last two columns show the distances derived from
Hipparcos parallaxes (van Leeuwen,  2007) (col. 7) and taken from other
sources (col. 8). For the former ones, only parallaxes whose uncertainty
$\sigma(\pi)$ is such that $\sigma(\pi)/\pi\leq 0.15$ have been considered. In
subsection \ref{Subsection:Distances} we discuss in detail the values of our
estimation of distances --actually, as we will see, {\it upper limits} in some
cases-- compared with other determinations.

From the Y$^2$ set, the tracks with $Z\!=\!0.02$ (solar) have been used for
those stars with metallicities [M/H] between $-0.10$ and $+0.10$. Assuming
that the relationship Fe/Fe$_\odot$=X/X$_\odot$ holds (in number of
particles), where X means any element with atomic number larger than 3, the
tracks corresponding to [M/H]=$+0.20$ would be those with $Z\!=\!0.03$. Since
these tracks are not computed in the original set, the results shown in Table
\ref{Table:MLAD} are the average of the values extracted using the tracks with
$Z\!=\!0.02$ and $Z\!=\!0.04$. For those stars with [M/H]=$+0.50$ the set
$Z\!=\!0.06$ was used. Finally for $\lambda$ Boo, we used the tracks with
$Z\!=\!0.0004$, which would correspond to [M/H]=$-1.7$. For HD 58647 and
BD+31 643 there are no high-resolution observations available to estimate
metallicities, therefore we assumed solar abundances to estimate their stellar
parameters; on the other hand, for HR 2174, 17 Sex and RR Tau the
metallicities found are solar or very close to it. The five objects have in
common that their locations in the HR diagram give masses above 5.2
$M_\odot$, which is the maximum in the Y$^2$ set, therefore, {\it only}
for these stars we used the Siess et al. tracks with solar metallicity.

{\renewcommand{\baselinestretch}{1.2}
\begin{table*}[th]
  \centering
  \caption[]{Stellar fluxes, masses, luminosities, ages and distances.}
  \begin{tabular}{lccrcrll}
    \hline\hline
Star & $F_*$ (W m$^{-2}$)& $M_*/M_\odot$ & $L_*/L_\odot$  & Age (Myr) & $d$ (pc) & $d_{\rm Hip}$ (pc) & $d_{\rm other}$ (pc)\\        
    \hline
HD 31648      & 3.308$\times 10^{-11}$ & 1.97$^{+0.04}_{+0.45}$ &  21.9$^{-7.8}_{+20.7}$   & 6.7$^{+1.0}_{-3.3}$  & 146$^{-29}_{+57}$  & 137$^{+31}_{-21}$ &                           \\
HD 58647      & 9.940$\times 10^{-11}$ & 5.99$^{-0.36}_{+0.38}$ & 911.2$^{-113.4}_{+221.2}$& 0.4$^{+0.1}_{-0.1}$  & 543$^{-35}_{+62}$  &                   &                           \\ 
HD 150193     & 2.825$\times 10^{-11}$ & 2.22$^{-0.04}_{+0.27}$ &  36.1$^{-9.2}_{+17.5}$   & 5.0$^{+0.5}_{-1.5}$  & 203$^{-28}_{+44}$  &                   &                           \\ 
HD 163296     & 6.583$\times 10^{-11}$ & 2.25$^{+0.11}_{+0.07}$ & 34.5$^{-5.1}_{+9.3}$     & 5.0$^{-0.3}_{-0.6}$  & 130$^{-10}_{+16}$  & 119$^{+12}_{-10}$ &                           \\ 
HD 179218     & 4.996$\times 10^{-11}$ & 2.56$^{-0.16}_{+0.21}$ &  63.1$^{-12.8}_{+17.0}$  & 3.3$^{+0.7}_{-0.7}$  & 201$^{-22}_{+25}$  &                   &                           \\ 
HD 190073     & 2.575$\times 10^{-11}$ & 5.05$^{-0.47}_{+0.54}$ & 470.8$^{-88.2}_{-189.2}$ & 0.6$^{+0.2}_{-0.2}$  & 767$^{-76}_{+139}$ &                   &                           \\ 
HR 10         & 1.737$\times 10^{-10}$ & 3.30$^{-0.37}_{+0.43}$ & 140.3$^{-41.9}_{+59.2}$  & 1.8$^{+0.5}_{-0.5}$  & 161$^{-26}_{+31}$  & 176$^{+24}_{-19}$ &                           \\ 
HR 2174       & 1.515$\times 10^{-10}$ & 6.82$^{-0.93}_{+1.02}$ &1012.7$^{-351.4}_{+217.2}$& 0.2$^{+0.1}_{-0.1}$  & 463$^{-89}_{+47}$  &                   &                           \\ 
HR 9043       & 8.784$\times 10^{-11}$ & 2.07$^{-0.21}_{+0.27}$ &  23.7$^{-6.7}_{+10.0}$   & 5.5$^{+2.0}_{-1.7}$  &  93$^{-14}_{+18}$  &  88$^{+4}_{-3}$   &                           \\ 
AS 442        & 9.753$\times 10^{-12}$ & 3.54$^{-0.37}_{+0.44}$ & 207.2$^{-60.1}_{+85.2}$  & 1.5$^{+0.5}_{-0.5}$  & 826$^{-130}_{+155}$&                   &                           \\ 
BD+31 643     & 2.483$\times 10^{-10}$ & 6.15$^{-0.29}_{+0.29}$ &1663.5$^{-234.9}_{+329.8}$& 0.4$^{+0.1}_{-0.1}$  & 464$^{-34}_{+36}$  &                   &                           \\ 
$\lambda$ Boo & 6.240$\times 10^{-10}$ & 1.66$^{-0.16}_{+0.19}$ &  19.1$^{-6.1}_{+9.0}$    & 2.8$^{+1.1}_{-0.8}$  &  31$^{-5}_{+7}$    &  30$\pm 1$        &                     \\ 
VX Cas        & 2.589$\times 10^{-12}$ & 2.33$^{-0.03}_{+0.14}$ &  30.8$^{-5.7}_{+5.8}$    & 6.4$^{+4.0}_{-1.5}$  & 619$^{-60}_{+58}$  &                   & 760$^{1,3}$               \\ 
SV Cep        & 3.390$\times 10^{-12}$ & 2.45$^{-0.08}_{+0.26}$ &  37.5$^{-5.1}_{+14.5}$   & 5.2$^{+1.4}_{-2.1}$  & 596$^{-43}_{+106}$ &                   & 440$^3$                   \\ 
24 CVn        & 4.686$\times 10^{-10}$ & 2.64$^{-0.11}_{+0.42}$ &  65.9$^{-19.1}_{+38.0}$  & 3.1$^{+0.6}_{-1.1}$  &  67$^{-11}_{+17}$  &  55$\pm 1$        &                           \\ 
51 Oph        & 4.991$\times 10^{-10}$ & 4.22$^{-0.30}_{+0.33}$ & 312.3$^{-62.3}_{+72.0}$  & 0.7$^{+0.4}_{-0.5}$  & 142$^{-15}_{+15}$  & 124$\pm 4$        &                           \\ 
T Ori         & 7.242$\times 10^{-12}$ & 2.42$^{+0.06}_{+0.17}$ &  50.2$^{-9.4}_{+17.4}$   & 4.0$^{-0.1}_{-0.8}$  & 472$^{-46}_{+76}$  &                   & 460$^1$                   \\ 
BF Ori        & 5.445$\times 10^{-12}$ & 2.61$^{-0.25}_{+0.41}$ &  61.6$^{-20.2}_{+32.4}$  & 3.2$^{+1.3}_{-0.9}$  & 603$^{-109}_{+142}$&                   & 450$^2$, 430$^3$          \\ 
UX Ori        & 4.414$\times 10^{-12}$ & 2.26$^{-0.25}_{+0.52}$ &  36.8$^{-13.7}_{+30.5}$  & 4.5$^{+2.0}_{-2.3}$  & 517$^{-108}_{+182}$&                   & 460$^1$, 400$^2$, 340$^3$ \\ 
V346 Ori      & 5.752$\times 10^{-12}$ & 2.52$^{-0.11}_{+0.31}$ &  61.4$^{-15.8}_{+27.3}$  & 3.5$^{+0.6}_{-1.1}$  & 586$^{-81}_{+118}$ &                   & 450$^2$\\ 
V350 Ori      & 1.741$\times 10^{-12}$ & 2.16$^{+0.13}_{+0.13}$ & 29.29$^{-5.8}_{+11.6}$   & 5.5$^{-0.1}_{-1.0}$  & 735$^{-77}_{+133}$ &                   & 460$^1$, 450$^2$          \\ 
XY Per        & 2.290$\times 10^{-11}$ & 2.79$^{-0.36}_{+0.59}$ &  85.6$^{-34.0}_{+64.3}$  & 2.5$^{+1.5}_{-0.8}$  & 347$^{-78}_{+190}$ &                   & 120$^1$, 350$^2$, 160$^3$ \\ 
VV Ser        & 2.869$\times 10^{-11}$ & 3.95$^{+0.15}_{+0.41}$ & 336.2$^{-89.4}_{+117.2}$ & 1.2$^{+0.1}_{-0.5}$  & 614$^{-88}_{+99}$  &                   & 440$^1$, 330$^3$, 230$^4$ \\ 
17 Sex        & 1.757$\times 10^{-10}$ & 6.35$^{-0.74}_{+0.80}$ & 940.5$^{-277.2}_{+295.8}$& 0.3$^{+0.1}_{-0.1}$  & 415$^{-30}_{+61}$  &                   &                           \\ 
RR Tau        & 5.673$\times 10^{-12}$ & 5.79$^{-0.94}_{+0.99}$ & 780.5$^{-351.9}_{+446.7}$& 0.4$^{+0.3}_{-0.2}$  &2103$^{-544}_{+535}$&                   & 800$^1$, 600$^2$          \\ 
WW Vul        & 3.315$\times 10^{-12}$ & 2.50$^{-0.06}_{+0.48}$ &  50.0$^{-19.11}_{+44.5}$ & 3.7$^{+3.7}_{-1.7}$  & 696$^{-149}_{+261}$&                   & 550$^1$, 534$^2$          \\\hline
HR 26 A       & 2.582$\times 10^{-10}$ & 3.08$^{-0.07}_{+0.07}$ & 106.9$^{-14.2}_{+15.7}$  & 175$^{-23}_{+14}$    & 115$^{-8}_{+8}$    &  94$\pm3$         &                           \\
              &                        & 3.01$^{+0.10}_{-0.01}$ & 104.0$^{-7.8}_{+11.9}$   & 2.6$^{-0.2}_{-0.1}$  & 114$^{-4}_{+6}$    &                   &                           \\
HR 4757 A     & 2.647$\times 10^{-9}$  & 2.74$^{-0.06}_{+0.07}$ &  69.0$^{-8.9}_{+9.7}$    & 260$^{-24}_{+14}$    &  29$^{-2}_{+2}$    &  27$\pm 1$        &                           \\ 
              &                        & 2.64$^{+0.04}_{+0.02}$ &  64.0$^{-4.2}_{+10.7}$   & 3.2$^{-0.1}_{-0.1}$  &  28$^{-1}_{+2}$    &                   &                           \\    
HR 5422 A     & 1.439$\times 10^{-10}$ & 2.66$^{-0.07}_{+0.08}$ &  60.4$^{-9.8}_{+11.0}$   & 267$^{-39}_{+23}$    & 116$^{-10}_{+10}$  & 110$\pm 5$        &                           \\ 
              &                        & 2.59$^{+0.09}_{+0.01}$ &  58.5$^{-5.7}_{+8.7}$    & 3.4$^{-0.2}_{-0.1}$  & 114$^{-6}_{+8}$    &                   &                           \\
49 Cet        & 2.064$\times 10^{-10}$ & 2.17$^{-0.03}_{+0.06}$ &  21.8$^{-1.0}_{+3.9}$    & 61$^{-46}_{+119}$    &  58$^{-1}_{+5}$    &  59$\pm 1$        &                           \\ 
              &                        & 2.15$^{-0.01}_{+0.11}$ &  21.6$^{-0.8}_{+4.4}$    & 8.9$^{+6.1}_{-2.4}$  &  58$^{-1}_{+6}$    &                   &                           \\
    \hline
\end{tabular}

\begin{minipage}{15cm} 
\underline{Notes to Table \ref{Table:MLAD}}: A discussion on the derivation of the distances given in col. 6 is done in 
subsection \ref{Subsection:Distances}. The distances given in col. 8 are taken from the works by $^1$Hern\'andez at al. (2004), 
$^2$Blondel \& Tjin A Djie (2006), $^3$Manoj et al. (2006) and $^4$Eiroa et al. (2008); see those works for the original references. For HR 26 A, 
HR 4757 A, HR 5422 A and 49 Cet we give the solutions derived using both Post-MS and PMS tracks, respectively 
(see subsection \ref{Subsection:Individual}).
\end{minipage}
\label{Table:MLAD}
\end{table*}
}

In Fig. \ref{Figure:hr} we show, at the left, the $\log g_*$ -- $\log T_{\rm
eff}$ HR diagram for the stars with metallicities between [M/H]=$-0.10$ and
$+0.10$, where tracks and isochrones for solar metallicity have been
superimposed, and at the right, the corresponding $\log L_*/L_\odot$ -- $\log
T_{\rm eff}$ HR diagram. The vertical error bars in the points $(\log T_{\rm
eff}, \log g_*)$ correspond to those listed in Table \ref{Table:Results}
whereas those in the points $(\log T_{\rm eff}, \log L_*)$ are the propagation
of the former from the first HR diagram to the second, hence their different
size up and down for a given object. A common horizontal error bar of
$\pm0.01$ dex has been assigned to the logarithm of effective temperatures,
which corresponds to $\pm 200$ K at $T_{\rm eff}$=10000 K.

\section{Discussion}
\label{Section:Discussion}

One important fact to bear in mind before starting the discussion is that the
results concerning the stellar mass, luminosity and age given in Table
\ref{Table:MLAD} {\it are distance-independent} determinations. Let us
remember that taking advantage of the knowledge of the stellar gravity we are
able to extract the mass and the age from a $\log g_*$ -- $\log T_{\rm eff}$
HR diagram, and then the luminosity from the corresponding point in the $\log
L_*/L_\odot$ -- $\log T_{\rm eff}$ HR diagram. 

In other works, e.g. those from Gerbaldi et al. (2001), Hern\'andez et
al. (2004) and Manoj et al. (2006), the authors extract these parameters from
a $\log L_*/L_\odot$ -- $\log T_{\rm eff}$ HR diagram, which implies a
knowlegde of the distance to the star, this being quite an uncertain parameter
in many occasions. Also, in the three works mentioned above, solar metallicity
is assumed for all the stars, whereas in our case, specific tracks for each
metallicity are used according to the abundance estimated for each object. The
different approaches of both formalisms makes it difficult a comparison among
the values of $M_*/M_\odot$, $L_*/L_\odot$ and age from our work and those
from other authors. That is the reason why in Table \ref{Table:MLAD} we only
provide for comparison values of distances from other works, which are
discussed later (see subsection \ref{Subsection:Distances}), whereas we do not
include any value of masses, luminosities and ages other than those found in
this paper. A detailed comparison of results for these parameters would be
lengthy and cumbersome.

\subsection{General considerations on the effective temperatures}
\label{Subsection:Temperatures}

The whole exercise carried out in this paper can be considered as a work of
basic astrophysics, in the sense that it concerns the determination of
fundamental parameters of a particular class of stars. The process has been
far from simple, due to the very nature of the objects studied, i.e. most of
them are not `normal' stars in the sense that their environment has a clear
influence on the photometric and spectroscopic properties. 

As we mentioned, the correct determination of the effective temperature, the
only initial parameter required to start the whole process, is a key problem
of this study. The calculation of $T_{\rm eff}$ hinges on the fit of the SED,
using synthetic models, from short wavelengths (optical or UV) to a wavelength
in the infrared, $\lambda_0$, where a change in the slope indicates that the
contribution from the disk starts to be noticeable. The photospheric part of
the SED can be affected by circumstellar (and interstellar) extinction, disk
occultation and, in addition, there can be non-photospheric contributions in
the UV (e.g. emission from accretion shocks or hot winds).  In the
determination of the effective temperatures carried out both by Mer\'{\i}n
(2004) and in this work, done by fitting Kurucz models to the observed SEDs
and given in Table \ref{Table:Results}, it was assumed that the spectral
energy distribution from the UV to $\lambda_0$, originates {\it entirely} in
the stellar photosphere. This hypothesis underlies the whole procedure and
should be critically assessed in each case\footnote{Since the UV
observations are not simultaneous with the optical and near-IR photometry, we
cannot estimate the non-photospheric contributions to the total UV flux;
however, even if these represent a $\sim\!10$\% of that flux, the results
would not differ very much, to within the uncertainties, from those under the
working hypothesis we use.}. In subsection \ref{Subsection:Individual} we
analyse in detail --among other stars-- VV Ser and RR Tau, for which the
determination of the effective temperature is especially problematic.

\subsection{Comparison with other abundance determinations}

Apart from the values of [M/H] around +0.80, for 24 CVn and XY Per, +0.50
for WW Vul, and the already well-known low metallicity of $\lambda$ Boo, the
remaining values of the abundances found in this work are solar or fairly
close to it. In addition to the stars analysed in this paper, the two HAeBe
stars studied by our group (Mer\'{\i}n et al. 2004), namely HD 34282 and HD
141569, show metallicities lower than solar, $-0.80$ and $-0.50$
respectively. No trend in the behaviour of the metallicities is observed in
the results for our sample.

As we mentioned in the Introduction, Acke \& Waelkens (2004) carried
out an abundance study of 24 Herbig Ae/Be and Vega-type stars searching for
the $\lambda$ Bootis phenomenon. For 20 of these stars the projected
rotational velocities, $v \sin i$, are less than 100 km/s, therefore the
authors were able to determine elemental abundances, instead of average values
as in our case. Stellar gravities were not estimated in that paper.  In order
to compare their results with ours we have computed a `mean' metal abundance
from those individual values. Details on how this was done are given in
Appendix A.

Guimar\~aes et al. (2006), as a result of a study of the accretion in Herbig
Ae/Be stars, also determined stellar gravities and metal abundances for 12
stars. The methodology of their work is almost identical to the one we have
used in this paper and also in the analysis of HD 34282 and HD 141569
(Mer\'{\i}n et al. 2004). They provide results for [Fe/H]. Apart from the
extreme Fe abundance of HD 100546 (not studied in this work), namely $-1.4$,
in agreement with the value $-1.30$ given by Acke \& Waelkens (2004), the
abundances for their sample vary from $-1.0$ to $+0.5$.

The three samples are not large enough to draw any definitive statistical
conclusion, although it is apparent that both in our work and in Acke \&
Waelkens' (2004) the majority of the stars tend to lie around metallicities
close to the solar one. The highest metallicities found in any of the three
works are around +0.80 whereas the lowest ones are below $-1.0$.

There is only one star in our sample in common with Acke and Waelkens', namely
HD 31648 (they found [M/H]=$+0.33\pm 0.08$ and [Fe/H]=$+0.07\pm 0.10$), and
another one with Guimar\~aes et al., namely HD 163296 (their result is
[Fe/H]=$+0.50\pm 0.10$). Our results are [M/H]=$0.00\pm 0.05$ for HD 31648 and
[M/H]=$+0.20\pm 0.10$ for HD 163296, in both cases somewhat lower than the
determinations of those two works.

\subsection{The estimation of the distances}
\label{Subsection:Distances}

Since the estimation of the distances hinges first on the determination of the
point ($\log T_{\rm eff}$, $\log g_*$) and then in the `translation' of this
point to a value of ($\log T_{\rm eff}$, $\log L_*/L_\odot$), any error in the
derivation of the effective temperature or the gravity would propagate to the
final value of the distance.

\begin{figure}
\centering
\includegraphics[width=9cm]{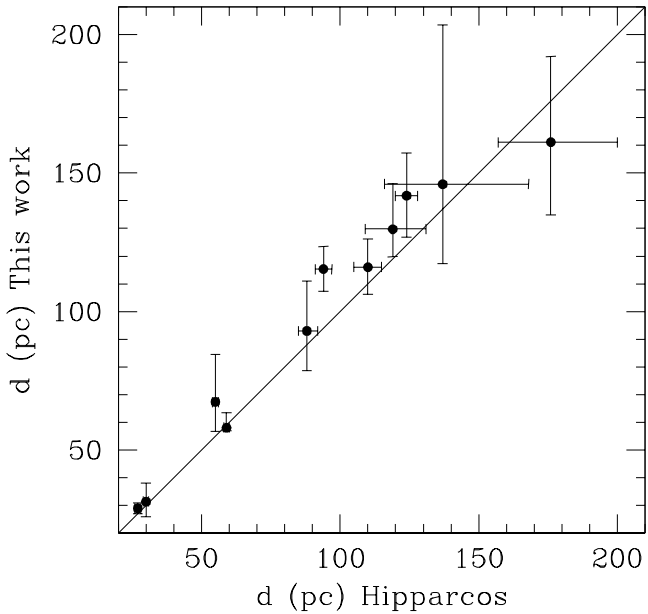}
 \caption{Plot showing the Hipparcos distances (x-axis) for the 11 stars in
 the sample with reliable parallaxes (see text for details), plotted against
 the corresponding distances estimated in this work (y-axis).}
 \label{Figure:distances}
\end{figure}

As it can be seen in Table \ref{Table:MLAD} there are 11 stars for which
Hipparcos parallaxes such that $\sigma(\pi)/\pi\leq 0.15$ --i.e. implying
reliable distances-- are available. The comparison between our estimations and
the Hipparcos-derived distances, illustrated in Fig. \ref{Figure:distances},
shows that in all those cases both determinations agree to within the
uncertainties. This is an additional indication of the internal coherence of
the overall results.

The remaining stars in Table \ref{Table:MLAD} either do not have Hipparcos
measurements or if they do, the large uncertainties make them
unusable. Instead, their distances have been estimated using a variety of
methods; in most cases the numbers found in the literature are just the
distances to the clouds where they could be embedded, which are, obviously,
rough approximations. In col. 8 of the Table we give results from the
compilations made by Hern\'andez et al. (2004), Blondel \& Tjin A Djie (2006)
and Manoj et al. (2006). It can be seen that in some cases those distances are
comparable to our results (e.g. T Ori, UX Ori, XY Per, WW Vul) but for other
stars the discrepancies are large. We can see a systematic trend between our
results (col. 6) and those from other authors (col. 8): out of 11 cases, in
nine of them the distances derived in this work are larger than the distance
to the parent cloud; the case of RR Tau is especially noticeable.

After exploring what might be the origin of this, we think that there is an
important fact that should be taken into account in interpreting these
discordant results. Note that to derive the distance $d$, we have used the
well-known expression $L_*\!=\!4\pi d^2 F_*$, where $L_*$ is the stellar
luminosity and $F_*$ is the observed photospheric flux, corrected for
reddening. To provide the correct distance, that simple expression must
contain the total flux, i.e. it is implicitly assumed that the {\it whole}
stellar photosphere is visible to the observer. If this does not occur, $F_*$
would be underestimated and, since the stellar luminosity has been fixed, the
implied distance would be larger than the real one. It could happen that, if
the inclination of the circumstellar disk is high or the opacity of part of
its material is very large (or both), the disk would block a fraction $\alpha$
of the stellar light, allowing only a fraction $F_*=(1-\alpha) F_*({\rm
total})$ to reach the observer. Since the material of the disk can be very
inhomogeneous, $\alpha$ would parameterize a sort of mean `filling factor' for
the occultation and obscuration of the stellar light caused by extremely high
values of the extinction (note that the observed flux $F_*$ can still be
affected by circumstellar and interestellar extinction). In addition, as we
pointed out in \ref{Subsection:Teffective}, the SED is built taking the
brightest photometry {\it among the available observations}, which could not
necessarily correspond to the true maximum brightness of the star: if that is
the case, the photospheric flux would be also less than the real one.

These arguments would put in agreement the discrepancies we mentioned
above. In subsection \ref{Subsection:Individual} we discuss in detail the
cases of VV Ser and RR Tau, and we see how this effect could have some impact
in our calculations.

The bottom line of this discussion is that the distances derived in this work
and given in col. 6 of Table \ref{Table:MLAD} should be used with caution.  In
particular, for those objects with high values of the extinction or high
inclination angles of their disks (e.g. almost edge-on disks, as it happens in
the UXORs), {\it they must be taken as  upper limits} of the actual distances.

\subsection{Notes on individual stars}
\label{Subsection:Individual}

\subsubsection{VV Ser}

This star is an intriguing HAeBe object.  The determinations of its parameters
found in the literature are discrepant, showing large differences in many
occasions. A summary of the results on the spectral type, luminosity class and
rotational velocities can be found in Table 1 of the paper by Ripepi et
al. (2007). Assignations such us A0 V, A3 II, A2 III or B6 to the spectral
type and values of $v \sin i$ from 85 to 229 km/s can be found (see references
in the paper by Ripepi et al.). Concerning the effective temperature, a wide
range from $9000\pm 1000$ K (Ripepi et al. 2007) to 13800 K (Hern\'andez et
al. 2004) is found\footnote{An even more extreme value of $\sim 7000$ K is
suggested by Ripepi et al. (2007). According to these authors, it allows to
reproduce the periodicities of the $p$-modes of the $\delta$ Scuti-like
pulsations observed in VV Ser.}.

\begin{figure*}
\centering
\includegraphics[width=16cm]{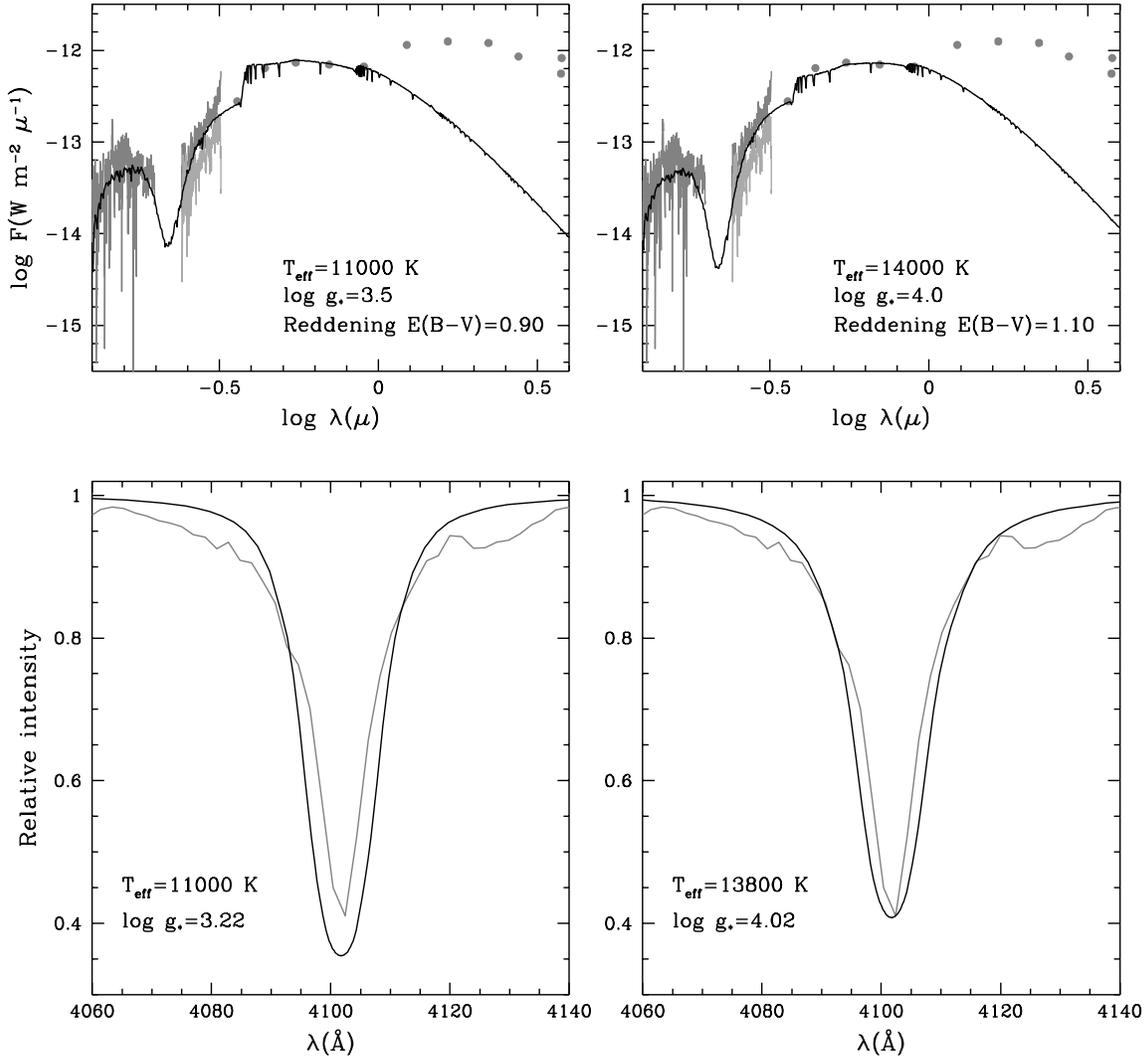}

 \caption{{\it Top}: The observed spectral energy distribution of VV Ser (grey
   lines and dots) and two fits to the photospheric contribution (black
   lines).  The UV spectra were obtained with the International Ultraviolet
   Explorer ({\it IUE}); the leftmost --and more noisy-- spectrum is SWP 48866
   (dark grey, obtained on 7/10/1993)) and the two spectra to its right are
   LWP 26520 (dark grey, 7/10/1993), and a merge of LWP 28921 and LWP 28931
   (light grey, 18--19/10/1994); these are the only spectra obtained by {\it
   IUE} of this object. The optical $UBVRI$ photometry was obtained on
   24/10/1998. Top left: The SED has been fitted by a Kurucz model with
   $T_{\rm eff}\!=\!11000$ K, $\log g_*\!=\!3.5$, reddened with
   $E(\!B-\!V)\!=\!0.90$.  Top right: The SED has been fitted by a Kurucz
   model with $T_{\rm eff}\!=\!14000$ K, $\log g_*\!=\!4.0$, reddened with
   $E(\!B-\!V)\!=\!1.10$. {\it Bottom}: The observed H$\delta$ line of VV Ser
   (grey line) and two fits with synthetic models (black lines). Bottom left:
   the best fit achieved using $T_{\rm eff}\!=\!11000$ K and $\log
   g_*\!=\!3.22$. Bottom right: the best fit for $T_{\rm eff}\!=\!13800$ K and
   $\log g_*\!=\!4.02$.}
 \label{Figure:vvser}
\end{figure*}

The estimates of the distance to the Serpens molecular cloud, where there
would be a possibility that the star is embedded into, have also many
uncertainties. The most recent studies converge to a somewhat shorter
distance: $225\pm 55$ pc for the front edge of the particular region of the
cloud studied (south of the cloud core) and a depth of 80 pc for the cloud
complex (Strai{\v z}ys et al. 2003). Eiroa et al. (2008) assign a distance of
$\sim\!230\pm 20$ to the Serpens cloud (see that reference for a complete
review on the properties of this region). There are no direct determinations
of the distance to VV Ser {\it itself}, and the values found in the literature
are linked to the distance to the cloud.

The initial value of $T_{\rm eff}$ for VV Ser we took to compute the stellar
parameters, according to the scheme presented in Section
\ref{Section:Computation}, was 11000 K, obtained by Mer\'{\i}n (2004), which
is in between the available determinations. The comparison of the synthetic
Balmer profiles with the observed ones led to a value of $\log g_*=3.22\pm
0.12$, which implied the following set of parameters:
$M_*\!=\!7.22^{-0.95}_{+1.00}$ $M_\odot$, $L_*\!=\!1638^{-467}_{+261}$
$L_\odot$, Age=0.2$^{+0.1}_{-0.1}$ Myr and, after combining the luminosity and
the observed flux, an upper limit to the distance $d=2224^{-344}_{+170}$ pc.

In Fig. \ref{Figure:vvser} (top) we show the observed SED of VV Ser from the
ultraviolet to the infrared, along with two fits to the photospheric part,
which covers the interval between 1200 \AA{} and 9000 \AA{} ($I$ band); from
this wavelength onwards, the contribution from the disk dominates. Note that
the ultraviolet spectra, obtained with the {\it IUE} observatory, are not
simultaneous with the optical and near infrared photometry (see caption of
Fig. \ref{Figure:vvser} for the dates); to our knowledge, unfortunately there
is not optical photometry available for the dates when the {\it IUE}
observations were taken, therefore it was not possible to compare the
brightness of the star in the two epochs in order to scale the data. In the
graph on the top left we can see how the SED is fitted reasonably well with a
Kurucz model\footnote{The fits to the SED shown in Fig. \ref{Figure:vvser}
(and later in Fig. \ref{Figure:rrtau}) have been done with models already
existing in the grids provided by Kurucz. There is a negligible difference
between the models with $T_{\rm eff}\!=\!11000$ K and $\log g_*\!=\!3.0, 3.5$,
therefore, any of the two models is, at the level of accuracy we can reach in
these fits, indistinguishable from the model with $\log g_*\!=\!3.22$. The
same argument holds for the models with $T_{\rm eff}\!=\!14000$ K and $\log
g_*\!=\!4.0$, and $T_{\rm eff}\!=\!13800$ K and $\log g_*\!=\!4.02$.} with
$T_{\rm eff}\!=\!11000$ K and $\log g\!=\!3.5$, reddened with
$E(\!B-\!V)\!=\!0.9$.

If this solution were correct, it would imply too young an age for the star
and disk incompatible with its apparent evolutionary status, according to the
scenario proposed by Malfait et al. (1998) (see Section 5 and Fig. 3, both of
that paper, for further details). Note that, according to the arguments
presented in \ref{Subsection:Distances}, the value $d=2224$ pc is just an
upper limit to the distance, and therefore, the remaining set of parameters
found cannot, in principle, be immediatly ruled out.  If the star were
actually placed at 250 pc, a simple calculation shows that $\alpha\simeq
0.99$, i.e., if the disk would block completely such a large fraction of the
photospheric light, this solution would be still coherent. This object was
classified as an UXOR star (Herbst \& Shevchenko, 1999; Rodgers, 2003), which
implies, according to the model proposed by Grinin (1988) and Grinin et
al. (1991), a moderate-to-high value for the inclination angle of the disk. In
the work by Pontoppidan et al. (2007a) a best-fitting model for the disk of VV
Ser is presented, giving a value $i=71.5\deg$ and a estimated range for this
parameter 65--$75\deg$.  Also, the {\it Spitzer} images of VV Ser and its
surroundings shown by Pontoppidan et al. (2007b) suggest an almost edge-on
position for the disk.

It is not strange that the effective temperature for this star be an elusive
parameter. Fig. \ref{Figure:vvser} (graph on the top right) shows a fit to the
SED using a Kurucz model computed with $T_{\rm eff}\!=\!14000$ K, close to the
value proposed by Hern\'andez et al. (2004), namely 13800 K, and $\log
g_*\!=\!4.0$. The Kurucz model has been reddened with $E(\!B-\!V)\!=\!1.10$.
This fit seems also reasonable, and the question is whether there is a way to
decide which one of the two solutions, if any, is correct.

In this case, the solution with a higher effective temperature seems to be
likely the correct one. The comparison of the shapes of the observed and
synthetic Balmer profiles for $T_{\rm eff}\!=\!11000$ K, under the assumption
that the star has a normal photosphere --which is implicit for all the stars
in this work-- did not result in a totally satisfactory fit. The shape of the
Balmer profiles hints towards a larger effective
temperature. Fig. \ref{Figure:vvser} (bottom) shows the H$\delta$ observed
profile and the best fits for $T_{\rm eff}\!=\!11000$ K and 13800 K. The
latter temperature provides a better global fit between the observed and
synthetic Balmer profiles, and in addition, leads to a set of final
parameters, shown in Tables \ref{Table:Results} and \ref{Table:MLAD}, which is
more coherent with the apparent evolutionary status of the star; in
particular, it sets an upper limit to the distance at 614$^{-88}_{+99}$ pc.
In this case, assuming again a distance of 250 pc to the star, the
fraction of photospheric light blocked would be $\alpha=0.83$.  We have
adopted the solution with the higher temperature as the best we can currently
deduce from the available data, although given the complexity of the system, 
the issue is still open.

\begin{figure}
\includegraphics[width=9cm]{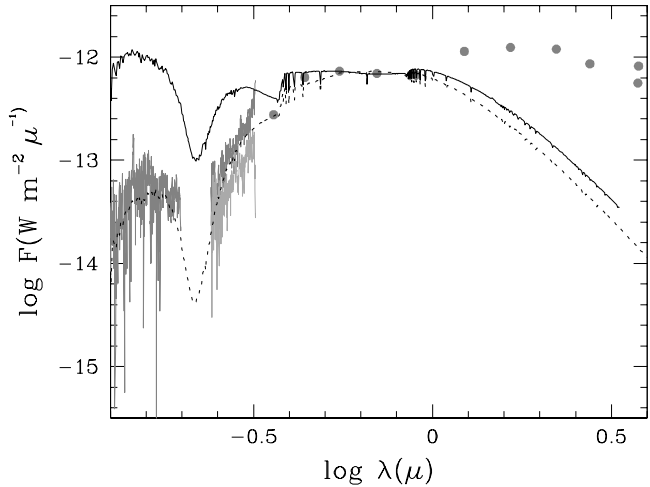}
 \caption{The observed spectral energy distribution of VV Ser (grey lines and
 dots) and two fits to its photospheric SED. The black dotted line (shown as a
 solid line in the upper-right panel of Fig. \ref{Figure:vvser}) corresponds
 to a Kurucz model with $T_{\rm eff}\!=\!14000$ K, $\log g_*\!=\!4.0$ reddened
 with the extinction law with $R_V\!=\!3.1$ and $E(B-V)=1.10$. The black solid
 line is the same Kurucz model reddened with with $E(B-V)=0.80$ but using the
 extinction law with $R_V\!=\!5.0$ (see text for details).}
 \label{Figure:vvser_extinction}
\end{figure}

VV Ser is the star in the sample for which a larger amount of
extinction was needed to reasonably fit the photospheric SED. In
subsection \ref{Subsection:Teffective} we mentioned that the
total-to-selective extinction $R_V\!=\!3.1$ was used instead of a
larger value ($R_V\!\sim\!5$) suggested by some studies that could be
more appropriate to treat the environment of HAeBe stars (Hern\'andez et
al. 2004). Figure \ref{Figure:vvser_extinction} shows the SED of VV
Ser plotted in the same way as in Fig. \ref{Figure:vvser} (upper
panels). The black dotted line represents the Kurucz model with
$T_{\rm eff}\!=\!14000$ K, $\log g_*\!=\!4.0$ reddened with the
extinction law with $R_V\!=\!3.1$ and $E(B-V)=1.10$ (also shown in
Fig. \ref{Figure:vvser}, upper-right panel) whereas the black solid
line represents the same Kurucz model, reddened with $E(B-V)=0.80$ but
using the extinction law with $R_V\!=\!5.0$; the parametrization given
by Cardelli et al. (1989) has been used. Whereas the results of the
SED fitting are roughly comparable and acceptable for both values of
$R_V$ for the optical and near-IR parts of the SED, the ultraviolet
part cannot be fitted with the $R_V\!=\!5.0$ law; if one attempts to
fit the ultraviolet part, the amount of extinction needed is such that
it is impossible to fit at the same time the UV and the optical and
near-IR intervals. Since the extinction law with $R_V\!=\!3.1$ allows
us to treat consistently the range of wavelengths from UV to the
near-IR, that value of $R_V$ was considered the most appropriate for
the analysis. The extinctions needed to fit the SEDs of the remaining
stars are lower than for VV Ser, therefore we decided to use the same
law for all the objects, giving an homogeneous treatment to the whole
sample.

\subsubsection{RR Tau}

This object is a highly variable HAeBe star also classified as UXOR-type
(Th\'e et al. 1994; Oudmaijer et al. 2001; Rodgers et al. 2002). The
determination of its absolute parameters is very difficult because the
variability observed does not allow to discern clearly what properties remain
unchanged, and hence can be attributable to the --non-variable-- stellar
photosphere, and what are totally due to the presence of circumstellar
material passing in front of the star.

\begin{figure}
\includegraphics[width=9cm]{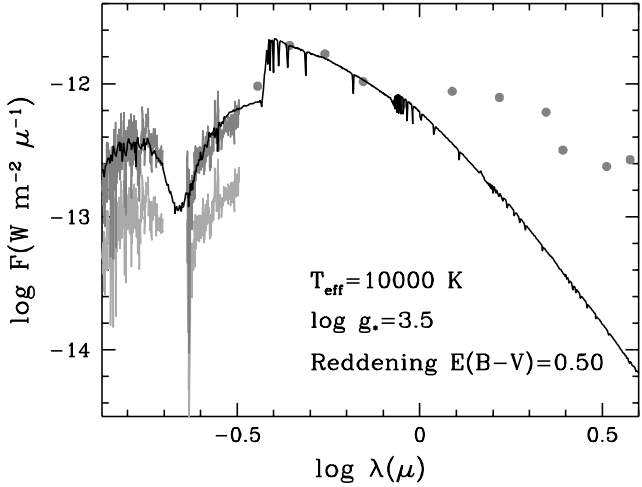}
 \caption{The observed spectral energy distribution of RR Tau (grey lines
   and dots) and the fit to the photospheric contribution (black line).  The
   UV spectra were obtained with the International Ultraviolet Explorer ({\it
   IUE}); in light grey lines the spectra SWP 48866 (7/10/1993)) and a merge 
   of LWP 26522 and LWP 26254 (7--8/10/1993) are plotted. Taking
   into account the value of the $V$ magnitude in the epochs when the
   ultraviolet observations were obtained ($V\!=\!12.2$), and when the
   $UBVRI$ photometry (dark grey dots) was taken ($V\!=\!10.92$, 31/01/1999),
   the {\it IUE} spectra have been shifted upwards by 0.512 dex (dark grey
   lines) in order to make `comparable' the observing conditions of the two
   epochs. The SED is fitted with a Kurucz model with $T_{\rm eff}\!=\!10000$ K,
   $\log g_*\!=\!3.5$, reddened with $E(\!B-\!V)\!=\!0.50$.}
 \label{Figure:rrtau}
\end{figure}

Fig. \ref{Figure:rrtau} shows the SED of this system from the ultraviolet to
the infrared, plotted in the same way as we did for VV Ser in the previous
paragraph. Given the high degree of variability of this star and the different
epochs when the ultraviolet {\it IUE} spectra and the optical photometry were
obtained, it is not trivial to compare both sets of observations. A simple
way, accurate enough for the purpose of this analysis, is to scale the
ultraviolet observations to the {\it UBVRI} points according to the value of
{\it V} in both epochs. The ultraviolet spectra shown in the figure were
obtained between JD 2449268.446 and 2449268.622 (mean JD 2449268.534). We have
found in the AAVSO archives that $V\!=\!12.3$ on JD 2449268.4 and $V\!=\!11.8$
on JD 2449269.4, with the brightness increasing the adjacent two days. An
interpolation between the magnitudes on the mentioned dates and JD 2449268.534
gives $V\!=\!12.2$ for the moment when the {\it IUE} observations were
obtained. Since $V\!=\!10.92$ when the optical photometry was taken, we have
shifted the ultraviolet spectra upwards by 0.512 dex to make comparable both
sets of observations\footnote{Note that this scaling is {\it grey}, in the
sense that we have moved upwards the ultraviolet spectrum by the same factor
for all wavelengths. The correct approach would be to take into account the
wavelength dependence of the extinction, which would imply also a
wavelength-dependent scaling factor.}. The best fit to this SED is obtained
with a Kurucz model with $T_{\rm eff}\!=\!10000$ K, $\log g_*\!=\!3.5$,
reddened with $E(\!B-\!V)\!=\!0.50$.

The value of the gravity deduced for this effective temperature by using the
Balmer profiles is $\log g_*\!=\!3.31\pm 0.13$, which implies the following
set of parameters: $M_*\!=\!5.79^{-0.94}_{+0.99}$ $M_\odot$,
$L_*\!=\!780.5^{-351.9}_{+446.7}$ $L_\odot$, Age=0.4$^{+0.3}_{-0.2}$ Myr and,
after combining the luminosity and the observed flux, a distance
$d\!=\!2103^{-544}_{+535}$ pc. Note that the age is fairly low to be in
agreement with the evolutionary scenario desribed by Malfait et
al. (1998). The value of the distance implied by this solution seems to be
quite large compared with estimates done for this star, which place the object
at around 800 pc (Finkenzeller \& Mundt, 1984). As it happened with the low
effective temperature case for VV Ser, the solution found for RR Tau could be
made compatible with that value of the distance if a fraction
$\alpha\!=\!0.86$ of the stellar light were completely blocked by the
disk. The classification of RR Tau as an UXOR star would favour a
moderate-to-high inclination for the disk, justifying this assumption.

Looking for a solution with a higher effective temperature, we have explored
the accuracy of the fits between the SED and the models of a grid with
effective temperatures from 11000 to 14000 K (steps of 1000 K), reddened with
colour excesses $E(B\!-\!V)$ from 0.50 to 0.90. In opposition to what happened
in the case of VV Ser, we have not found a solution that fits reasonably well
the observed SED, therefore we have adopted for RR Tau the effective
temperature, gravity and stellar parameters shown in Tables
\ref{Table:Results} and \ref{Table:MLAD}. Our results are consistent with
those reported by Grinin et al.  (2001; $T_{\rm eff}\!=\!9750$ K, $\log
g_*\!=\!3.5$, solar metalicity) and Rodgers et al. (2002; 9000 $\leq T_{\rm
eff} \leq$ 10000, 3.0 $\leq \log g_* \leq$ 3.5).

\subsubsection{HR 26 A, HR 4757 A and HR 5422 A}

These three stars were included in the sample as hot companions of post-T
Tauri stars, extracted from the Lindroos catalogue (Lindroos, 1985).  Gerbaldi
et al. (2001) carried out a systematic study of a sample binary systems with
post-T Tauri secondaries. Both components, A and B, of HR 26 (HD 560), HR 4757
(HD 108767) and HR 5422 (HD 127304) were analysed in detail in that work and
effective temperatures, gravities, luminosities, masses and ages were
computed. The effective temperatures and gravities were obtained by those
authors from calibrations using both $ubvy\beta$ and Geneva photometry; the
luminosities, masses and ages were estimated in a different way from that
followed in this paper: using the Hipparcos parallax and the $V$ magnitude of
each star, they computed $M_{\rm bol}$ and then $L_*/L_\odot$, placing the
star in a $\log L_*/L_\odot - \log T_{\rm eff}$ HR diagram and from its
position, the remaining parameters were derived.

{\renewcommand{\baselinestretch}{1.2}
\begin{table}[h]
  \caption[]{Some stellar parameters for HR 26, HR 4757 and HR 5422.}
  \begin{tabular}{lcccc}
    \hline\hline
Star      & $T_{\rm eff}$ (K)  &     $\log g_*$    &     Age (Myr)  & Tracks  \\
\hline
HR 26 A   &    11400           &   $4.14\pm 0.04$  &   $130^{-40}_{+47}$, $140^{-30}_{+64}$   &  1,2      \\
HR 26 B   &                    &                   &   $20^{+13}_{-8}$, $30^{+70}_{-8}$       &  1,2      \\
HR 4757 A &    10400           &   $4.05\pm 0.06$  &   $200^{-30}_{+40}$, $200\pm 30$         &  1,2      \\  
HR 4757 B &                    &                   &   $40_{-17}$, $90_{-50}$                 &  3,4      \\
HR 5422 A &    10040           &   $4.12\pm 0.02$  &   $240^{-30}_{+60}$, $240^{-20}_{+60}$   &  3,4      \\
HR 5422 B &                    &                   &   $50_{-20}$, $100_{-50}$                &  3,4      \\
\hline
HR 26 A   &    11400           &   $4.08\pm 0.05$  &   175$^{-23}_{+14}$   &   5                          \\
          &                    &                   &   1.8$^{+0.4}_{-0.2}$ &   6                          \\ 
HR 4757 A &    10400           &   $4.06\pm 0.05$  &   260$^{-24}_{+14}$   &   5                          \\
          &                    &                   &   2.5$^{+0.5}_{-0.2}$ &   6                          \\
HR 5422 A &    10250           &   $4.08\pm 0.06$  &   267$^{-39}_{+23}$   &   5                          \\
          &                    &                   &   2.1$^{+0.1}_{-0.4}$ &   6                          \\
\hline
\end{tabular}

\begin{minipage}{8.5cm}
\underline{Notes to Table \ref{Table:hptt}}: The results shown in the upper
part of the table are from Gerbaldi et al. (2001), those in the lower part are
derived in this work. The tracks used for computing the ages are: $^1$Schaller
et al. (1993), $^2$Girardi et al. (2000), $^3$D'Antona \& Mazzitelli,
I. (1998), $^4$Palla \& Stahler (1999), $^5$Y$^2$ post-MS, $^6$Y$^2$ PMS.
\end{minipage}
\label{Table:hptt}
\end{table}
}

\begin{figure*}[t]
\centering
\includegraphics[width=18cm]{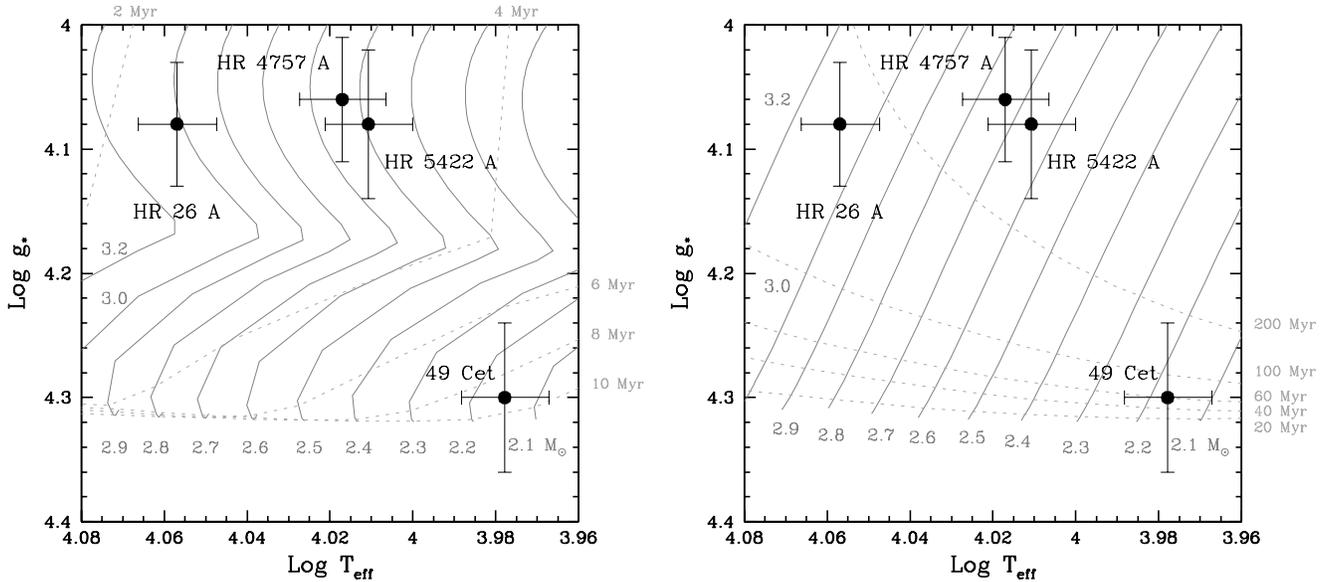}
\caption{Two HR diagrams $\log g_* - \log T_{\rm eff}$ where the positions of
the stars HR 26 A, HR 4757 A, HR 5422 A and 49 Cet have been plotted. The
error bars in $\log T_{\rm eff}$ correspond to uncertainties of $\pm 250$
K. The diagrams show the stars plotted on PMS (left) and post-MS (right)
tracks (solid lines) and isochrones (dashed lines) for solar metallicity from
Yi et al. (2001).}
\label{Figure:4stars}
\end{figure*}

Table \ref{Table:hptt} shows a comparison of the results from Gerbaldi et al.'
work and ours. The results shown in the upper part are extracted from Tables 1
(effective temperatures and gravities), 7 and 9 (ages of the early and
late-type stars, respectively) of Gerbaldi et al. (2001); those in the lower
part are from Tables \ref{Table:Results} and \ref{Table:MLAD} of this
work. For the late-type stars we just list the ages found by Gerbaldi et
al. (2001) in order to have all data needed for the discussion of the results.

One should not be confused with the terminology `hot companion of a post-T
Tauri star' attributed to these stars in the sense of assuming as a real fact
that they are physically linked to the late-type companion. Gerbaldi et
al. (2001) came to the conclusion that for HR 26, the ages of the components
were compatible with a physical association between them due to the overlap of
the error bars of the ages. The lack of that overlap lead these authors to
conclude that the components of HR 4757 and HR 5422 are not linked (see Table
\ref{Table:hptt}). Note, however, that the result on HR 26 hinges on a 
large error in the determination of the age of the late-type star,
therefore, in our opinion, that claim is very uncertain.

Strictly speaking, only with the data corresponding to the effective
temperature and gravity (or luminosity), it is impossible to say whether these
three stars are in a PMS or a post-MS evolutionary stage. When the
corresponding points ($\log T_{\rm eff}$, $\log g_*$) --or ($\log T_{\rm
eff}$, $\log L_*/L_\odot$), if the $V$ magnitudes, bolometric corrections and
Hipparcos distances are used-- are placed in the HR diagram, they lie above
the main sequence. In Figure \ref{Figure:4stars} we can see two HR diagrams, on
the left we have plotted our results on the Y$^2$ tracks and isochrones for
PMS stars and on the right, on the Y$^2$ set for the post-MS. In Tables
\ref{Table:MLAD} and \ref{Table:hptt} we give the results derived for both
cases. The PMS ages are small, less than 3 Myr in all cases, whereas the
post-MS ages are in the range 150-300 Myr, consistent with the results by
Gerbaldi et al., although slightly larger.

The SEDs do not show any remarkable infrared excess: HR 26 A and HR 5422 A do
not have {\it IRAS} detections, and HR 4757 A has {\it IRAS} fluxes at 12 and
25 $\mu$m lying slightly above the photospheric continuum and only upper
limits at 60 and 100 $\mu$m.  HR 26 A has a detection at 1.3 mm above the
photospheric continuum (Altenhoff et al., 1994). However, the absence of an
infrared excesses cannot, in principle, rule out the results extracted from
PMS tracks, since it is not guaranteed that {\it all} A-type PMS stars in the
age range 1-10 Myr have large infrared excesses indicating the presence of a
circumstellar disk. None of the stars show H$\alpha$ in emission.

Pallavicini et al. (1992) carried out optical spectroscopy of post-T Tauri
star candidates. HR 26 B (G5 Ve) and HR 4757 B (K2 Ve) show strong Ca {\sc
ii} emissions and Li absorptions, which are indications of youth, whereas HR
5422 B (K1 V) does not have Ca {\sc ii} emission and the Li absorption is
weak, which implies that this object must be older than the other two
stars. The ages derived by Gerbaldi et al. for the `B' components, listed in
Table \ref{Table:hptt}, are consistent with those observations, but in all
cases they are larger (under the assumption that the `A' components are PMS
objects) or smaller (assuming that the `A' components are post-MS stars), by
very significant differences, than the ages of the corresponding early-type
companions. Therefore, we can conclude that in none of the three cases the pairs
are physically linked.

\subsubsection{49 Cet and other stars without H$\alpha$ emission}

In Table \ref{Table:Sample} we indicate what stars show the H$\alpha$ line in
emission. It is well known that stars without disks with effective
temperatures in the range of those considered in this work show the H$\alpha$
line --and all the lines of the Balmer series-- in absorption. The presence of
emission in PMS stars with protoplanetary disks is an indication of winds
(Finkenzeller \& Mundt, 1984), accretion (e.g. Muzerolle et al. 2005), or a
combination of both (Kurosawa et al. 2006)\footnote{If the star is of a
spectral type later than F, the H$\alpha$ emission will have also a
chromospheric contribution}. Since the typical lifetime of protoplanetary
disks is a few million years (Alexander, 2007), these phenomena are an
indication of youth, and therefore the status of PMS assumed {\it a priori}
for all the stars in our sample with H$\alpha$ emission is coherent.

However, the fact that a star does not show H$\alpha$ in emission does not
imply, in principle, that the object has abandonned the PMS phase. In our
sample there are stars with no emission but showing remarkable infrared
excesses, indicative of the presence of a disk. 

The case of 49 Cet is especially interesting.  The peculiarity of this object
is that its disk has retained a substantial amount of molecular gas, which is
a typical feature of protoplanetary disks (i.e. ages less than $\sim\!10$
Myr) but shows dust properties similar to those of debris disks (Hughes et
al. 2008). The spectrum, as can be seen in Fig. \ref{Figure:panel}, is purely
photospheric, and does not show any signature of circumstellar activity. 

In Table \ref{Table:MLAD} we give the stellar parameters for this object using
both PMS and post-MS tracks and isochrones.  Fig. \ref{Figure:4stars} shows
the position of the star in two HR diagrams; on the left, the Y$^2$ PMS
tracks and isochrones have been used, on the right, the post-MS set has been
plotted. Thi et al. (2001) assigned an age of 7.8 Myr to 49 Cet using PMS
tracks, which is in good agreement with our PMS determination. Our estimations
of the effective temperature and gravity place the star almost on the main
sequence, therefore its evolutionary status can be that of a star either on
the PMS phase, entering the MS, or just evolving out of it. Actually, the
uncertainties in the gravity are such that they allow the star to be {\it
exactly} on the main sequence.

The region of the HR diagram where 49 Cet lies is particularly difficult to
carry out an estimation of ages: the isochrones are very close to each other
and therefore a small change in gravity implies a large change in age. In the
case of this star, the range of ages covered assuming that the star is in a
PMS phase or in a post-MS stage is 8.9--61 Myr, hence any age in between
these two determinations could be a reasonable result. Hughes et al. (2008)
suggest that the disk might be in an unusual transitional phase and our
results are in agreement with this hypothesis.

As it can be seen in Table \ref{Table:Sample}, in addition to HR 26 A, HR 4757
A and HR 5422 A, already studied in the previous subsection, there are other
stars without H$\alpha$ emission. Those objects have been classified as
A-shell or Vega-like stars, implying that the corresponding infrared excesses
are not very large. However, these stars deserve further and more detailed
study since the ages derived from PMS tracks (see Table \ref{Table:MLAD}) are
in all cases less than 6 Myr and some objects, e.g. HR 10 and $\lambda$ Boo show
a high degree of circumstellar activity.

\section{Final remarks}
\label{Section:Remarks}

Effective temperatures, gravities, metallicities, masses, luminosities, ages
and distances (or upper limits), have been obtained for a sample of 30
early-type objects, more precisely 27 HAeBe, Vega-type and Ash stars, and
three hot --suspected and finally ruled out-- companions of post-T Tauri
stars. The results obtained are described in detail in Section
\ref{Section:Results} and discussed in Section \ref{Section:Discussion}.

As a bottom line of this paper we would like to stress the difficulty of
finding the correct stellar parameters for this kind of objects. As we have
seen in the cases of VV Ser and RR Tau, shown in subsection
\ref{Subsection:Individual}, the spectral energy distributions are difficult
to interpret and the assumption that the ultraviolet, optical and near
infrared parts of the SED originate entirely in the photosphere must be
assessed in each case. Simultaneity of the observations in all wavelength
ranges is crucial to build a consistent SED; in this work we have used
ultraviolet data that were obtained long before the optical and near infrared
data, this must be always be present in the analysis of the SED. Also, the
wings of the Balmer lines used to compute the gravity are supposed to be
purely photospheric, without any external contamination. Note that these
hypotheses, underlying the whole procedure, are the starting points to
determine the remaining set of stellar parameters.  The extinction caused by
the environment and the variability of the objects can also mask the `true
photosphere', if that concept is applicable to this kind of stars.

On the other hand, the uncertainties in the determination of temperatures and
gravities put a limit to the accuracy of the ages estimated using
isochrones. As it can be seen in Figs. \ref{Figure:hr} and
\ref{Figure:4stars}, depending on the region of the HR diagram where a given
star lies, the shape and position of the isochrones provides in some cases a
wide range of ages. In particular, when the star lies very near the main
sequence, as it happens to 49 Cet, even the determination of the evolutionary
status as PMS or post-MS object can be a problem.

In view of these considerations we suggest to take the ages found in this paper
--and in general, all the determinations of ages at this evolutionary stage
using isochrones-- with caution, assessing each particular case in a critical
way and complementing their values with those of an observable quantity, such
as $F_{\rm IR}/F_*$, i.e. the ratio between the excess flux from the disk and
the photospheric flux, which can be taken as a proxy for the age of the
system.

We would also like to remark that the method proposed to obtain the stellar
parameters is distance-independent, in the sense that it is based on the
determination of the effective temperature and gravity, which are used, via a
transformation from a $\log g_* - \log T_{\rm eff}$ HR diagram into a $\log
L_*/L_\odot - \log T_{\rm eff}$ HR diagram, to compute the remaining
parameters. The method has proven to be reliable, given the agreement between
the distances found in this work and those available from Hipparcos
measurements (see subsection \ref{Subsection:Distances} and
Fig. \ref{Figure:distances}). For those stars with unknown distances, our
determinations are upper limits to the real values.

The tasks carried out in this paper, resulting in the characterization of the
stars, are necessary steps to study and model their circumstellar disks. Work
is already in progress to show a global picture of the behaviour of the disks
of these stars and their interaction, via accretion processes, with the star.

\begin{acknowledgements}
This work has been supported in part by a grant attached to the project
AYA2005--00954, funded by the Spanish Ministry of Science and Education.

The authors are grateful to Bram Acke for sending us some high resolution
spectra of HD 31648, Piercarlo Bonifacio for his help with the use of SYNTHE,
Robert Kurucz for some clarifications on his model atmospheres and Sukyoung Yi
for kindly providing unpublished smoothed evolutionary tracks and for advice
on their use. We would also like to thank the staff at the Calar Alto and
Roque de los Muchachos observatories for their support during the observing
runs.

We acknowledge with thanks the variable star observations from the AAVSO
International Database contributed by observers worldwide and used in this
research.

Finally, we also thank Jes\'us Hern\'andez, the referee, for the careful
reading of the original manuscript and many comments and suggestions.

\end{acknowledgements}

\section*{Appendix A}

In this Appendix we treat the problem of how to compute a meaningful `mean
metal abundance' from measurements of individual elemental abundances. Let us
consider the case of a set of elemental abundances, some of which have
been obtained using lines of two different ionization stages.

Let $x_i(j)$ be the elemental abundance of the element $i$ obtained using
absorption lines of its $j$-th ionization stage, i.e.  $x_i(j)=[{\rm
A}_i(j)/{\rm H}]$ and let $\Delta x_i(j)$ be the corresponding error.

In the particular case when an element has different measurements of the
abundances using lines of two ionization stages, namely I and II, we obtain
the abundance of that element by averaging the individual values weighted by a
factor equal to the inverse of the square of their errors:

\begin{equation}
x_i=[{\rm A}_i/{\rm H}] = \frac{x_i({\rm I})/\Delta^2 x_i({\rm I})
                               +x_i({\rm II})/\Delta^2 x_i({\rm II})}
                          {1/\Delta^2 x_i({\rm I})+1/\Delta^2 x_i({\rm II})}
\end{equation}

\noindent and

\begin{equation}
\Delta^2 x_i=\sigma^2=\frac{1}{1/\Delta^2 x_i({\rm I})+1/\Delta^2 x_i({\rm II})}
\end{equation}

In this way, if the measurement from one ionization state is much more precise
than the other, the average value and its error will approach the best
ones. However, if both measurements have equal precision, the average error will
be reduced by a factor $\sqrt{2}$.

Once we have all the elemental abundances and their errors, i.e. the pairs
$x_i$, $\Delta x_i$, we can proceed with the calculation of the mean metal
abundance:

Let $B_i$ be

\begin{equation}
B_i=\left(\frac{N_{\rm A_i}}{N_{\rm H}}\right)_\odot
\end{equation}

The mean metallicity can be computed as:

\begin{equation}
{\rm [M/H]}=\log \sum_i B_i\,10^{x_i} - \log \sum_i B_i
\label{MoverH}
\end{equation}

\noindent where the sums extended over all $i$ metallic elements.
Since only a few metallicities are measurable for each star, the sum
will be restricted to those atoms, assuming that the remaining unknown
abundances will not alter significantly the result.

The error in the determination of the mean abundance can be treated as
follows: since ${\rm [M/H]}=f(x_1,\dots,x_N)$ then

\begin{equation}
\sigma({\rm [M/H]}) =
       \sqrt{\sum_i\left(\frac{\partial f}{\partial x_i}
                        \Delta x_i\right)^2} 
\end{equation}

\noindent where

\begin{equation}
\frac{\partial f}{\partial x_i}=\frac{B_i\,10^{x_i}}{\sum_j B_j\,10^{x_j}}
\end{equation}

\noindent according to equation (\ref{MoverH}).

Acke \& Waelkens (2004) provide the values of [A$_i$/H] and use the
solar abundances by Anders \& Grevesse (1989), so in principle, it is
not difficult to compute an average abundance to be compared with ours. The
problem, however is a bit more subtle and a direct comparison is not as
trivial as it may appear.  Acke \& Waelkens (2004) divide the results
in three tables: their table 2 contains the C, N, O and S abundances; table 3
contains the abundances of `metals', where by `metals' they mean Mg, Si, Ca,
Ti, Fe and Sr, and finally table 4 contains the abundances of other elements,
namely Na, Al, Sc, V, Cr, Mn, Ni, Zn, Y, Zr and Ba.

\begin{table}[th]
  \caption[]{Mean abundances for the sample of Acke \& Waelkens (2004).}
  \begin{tabular}{lcc}
    \hline\hline
Star & Metals & [Fe/H]\\        
    \hline
HD 4881         &$-0.78\pm0.05$       & $-0.64\pm0.09$ \\
HD 6028         &$-0.18\pm0.06$       & $-0.13\pm0.01$ \\
HD 17081        &$-0.09\pm0.13$       & $-0.26\pm0.19$ \\
HD 17206        &$-0.07\pm0.05$       & $-0.09\pm0.11$ \\
HD 18256        &$+0.06\pm0.03$       & $+0.06\pm0.11$ \\
HD 20010        &$-0.17\pm0.05$       & $-0.16\pm0.12$ \\
HD 28978        &$\,0.00\pm0.08$      & $+0.18\pm0.11$ \\
HD 31293        &$-0.35\pm0.03$       & $-1.04\pm0.12$ \\ 
HD 31648        &$+0.33\pm0.08$       & $+0.07\pm0.10$ \\
HD 33564        &$+0.14\pm0.03$       & $-0.38\pm0.13$ \\
HD 36112        &$-0.02\pm0.07$       & $-0.14\pm0.08$ \\
HD 95418        &$+0.07\pm0.04$       & $+0.23\pm0.06$ \\
HD 97048        &$-0.75\pm0.06$       & $-$            \\
HD 97633        &$+0.08\pm0.08$       & $+0.01\pm0.08$ \\
HD 100453       &$+0.03\pm0.05$       & $-0.09\pm0.13$ \\
HD 100546       &$-1.05\pm0.08$       & $-1.30\pm0.13$ \\
HD 102647       &$-0.19\pm0.08$       & $-0.21\pm0.09$ \\
HD 104237       &$+0.06\pm0.05$       & $+0.09\pm0.12$ \\
HD 135344       &$+0.01\pm0.07$       & $-0.08\pm0.10$ \\
HD 139614       &$-0.32\pm0.04$       & $-0.27\pm0.11$ \\
HD 190073       &$+0.02\pm0.05$       & $-0.04\pm0.10$ \\
HD 244604       &$-$                  & $-0.25$        \\
HD 250550       &$-0.40\pm0.06$       & $-0.88\pm0.12$ \\
\hline
\end{tabular}
\label{Table:AW04}
\end{table}

Instead of computing the mean metallicity using {\sl all} the elemental
abundances, the criterion we have chosen is to include only abundances of the
`metals'. These averages, and their uncertainties, are shown in the second
column of Table \ref{Table:AW04}. The reason for this is that, since the
atomic fractions of C, N, O and S are much larger than those of the remaining
elements, these four elements would dominate the whole mean value. As the
optical spectrum of early-type stars is dominated by lines of `metals', the
abundances we have computed for the stars in our sample are, in fact, related
to to those species and not to the lighter elements, namely, C, N, O and S.

For comparison we give in the third column of Table \ref{Table:AW04} the Fe
abundances. Note that they can be quite different from the mean abundances
obtained from the metals. This comes from the fact that the elemental
abundance ratios in these stars differ in general from the solar ones, so the
abundances of other `metals', mainly Mg and Si, have a strong weight on the
mean abundance.

\end{document}